\documentclass[journal,romanappendices]{IEEEtran}
\usepackage{cite}
\usepackage[cmex10]{amsmath}%
\usepackage{amssymb}%
\usepackage{amsthm}%
\usepackage{amsfonts}
\usepackage{algorithmic}
\usepackage[pdftex]{graphicx}
\usepackage{textcomp}
\usepackage[T1]{fontenc}%
\usepackage{bm}
\usepackage{array}%
\usepackage{mdwmath}
\usepackage{mdwtab}
\usepackage[caption=false]{subfig}
\usepackage{url}
\usepackage{xcolor}

\providecommand{\U}[1]{\protect\rule{.1in}{.1in}}
\newtheorem{theorem}{Theorem}
\newtheorem*{theorem*}{Theorem}

\newcommand{\im}{j} 

\DeclareMathOperator{\trace}{Tr}

\newcommand{\ket}[1]{\left| #1 \right>} 
\newcommand{\bra}[1]{\left< #1 \right|} 
\mathchardef\mhyphen="2D 

\begin{document}

\newcommand{\papertitle}{Quantum-Enhanced Transmittance Sensing}
\title{\papertitle{}}

\author{Zihao Gong, \IEEEmembership{Student Member, IEEE}, Nathaniel Rodriguez, \IEEEmembership{Student Member, IEEE}, Christos N. Gagatsos, Saikat Guha, \IEEEmembership{Senior Member, IEEE}, and Boulat A. Bash, \IEEEmembership{Member, IEEE}
\thanks{This paper was presented in part at the 2021 International Symposium on Information Theory (ISIT 2021) \cite{gong21losssensing}.}
\thanks{ZG and BAB were supported by the National Science Foundation (NSF) under Grant No.~CCF-2045530. NR was supported by the NSF under Grant Nos.~EEC-1941583 and CCF-2006679. CNG and SG acknowledge funding support from the Office of Naval Research, grant number N00014-19-1-2189.}
\thanks{Zihao Gong and Nathaniel Rodriguez are with Electrical and Computer Engineering Department, The University of Arizona, Tucson, AZ 85721 USA (email: zihaogong@email.arizona.edu and nrod968@email.arizona.edu).}
\thanks{Christos N.~Gagatsos is with the College of Optical Sciences, The University of Arizona, Tucson, AZ 85721 USA (email: cgagatsos@arizona.edu).}
\thanks{Saikat Guha and Boulat A.~Bash are with the Electrical and Computer Engineering Department, The University of Arizona, Tucson, AZ 85721 USA, and also with the College of Optical Sciences, The University of Arizona, Tucson, AZ 85721 USA (email: saikat@optics.arizona.edu and boulat@arizona.edu).}
}

  \maketitle

\begin{abstract}
We consider the problem of estimating unknown transmittance $\theta$ of a target bathed in thermal background light.
As quantum estimation theory yields the fundamental limits, we employ the lossy thermal-noise bosonic channel model, which describes sensor-target interaction quantum mechanically in many practical active-illumination systems (e.g., using emissions at optical, microwave, or radio frequencies). 
We prove that quantum illumination using two-mode squeezed vacuum (TMSV) states asymptotically achieves minimal quantum Cram\'{e}r-Rao bound (CRB) over all quantum states (not necessarily Gaussian) in the limit of low transmitted power.  We characterize the optimal receiver structure for TMSV input, and show its advantage over other receivers using both analysis and Monte Carlo simulation.
\end{abstract}

\section{Introduction}
\label{sec:intro}

A precise measurement of power transmittance is a fundamental task in engineering.
It translates to measuring both target reflectance in active sensing systems, such as RADAR and LIDAR, and signal distortion from attenuation in communications systems.
Transmittance is also critical to quantum-system design.
It determines the precision of quantum methods for phase estimation \cite{escher11metrology,demkowicz12elusive}, the point-to-point quantum-communication capacity \cite{pirandola17QKDcap}, and whether a quantum channel preserves the entanglement \cite{namiki14ebcondition}.

The importance of measuring transmittance led to the development of classical signal processing methods covering many practical scenarios \cite{kay93statSP1, vantrees01part1}.
However, the fundamental precision limits for all sensing tasks as well as the approaches to achieve these limits are governed by quantum information theory \cite{helstrom76quantumdetect, nielsen00quantum, wilde16quantumit2ed, hayashi17qit}.
As we briefly discuss in Section \ref{sec:quantum_estimation}, this is because quantum information methods optimize the underlying physical measurement process that generates the data for the estimator, as well as the estimator itself. 
Indeed, quantum-enhanced sensing systems can significantly outperform those limited by classical methodology \cite{degen17quantumsensing, pirandola18quantumsensing}.

Consider active sensing of target reflectance, using optical, microwave, or radio-frequency emissions in the presence of background Gaussian noise.
This task is modeled quantum mechanically by the measurement of power transmittance of a lossy thermal-noise bosonic channel.
Despite the progress in quantum transmittance sensing \cite{monras07loss, adesso09optlossest, nair18loss, monras11tmsvopt,  nair20qi, jonsson22gaussianlosssensing, goldberg22multiparameterlossxanadu}, outlined briefly in Section \ref{sec:problemstatement}, a design of a sensor transceiver that attains the quantum limit in the presence of environmental thermal noise, has been elusive.
In fact, the authors of \cite{nair20qi}, upon establishing the fundamental lower bound on the mean squared error (MSE) of quantum transmittance estimation, question whether a thermal-noise scenario even exists where this bound is saturated.
Here, we answer this question affirmatively: the lower bound is achievable for probes with low transmitted power.
Furthermore, we characterize the corresponding transceiver and provide analysis, and simulations supporting its near-term physical implementation.

We begin by describing in Section \ref{sec:prerequisites} our notation and the lossy thermal-noise bosonic-channel model.
We then cover the basics of quantum estimation theory.
This allows us to introduce the quantum perspective on the transmittance-sensing problem and to consider the use of \emph{quantum illumination} (QI) \cite{lloyd08quantumillumination, tan08qigaussianstates, guha09quantumilluminationOPA, zhang15eesensing, sanz17estimationQI, shapiro20QIstory, nair20qi}, which, in general, improves precision by using entanglement between the transmitted probe and a reference state retained in the transceiver.
In Section \ref{sec:tmsv} we prove that probes constructed from two-mode squeezed vacuum (TMSV) states can achieve the ultimate bound in the limit of low transmitted power.
As was done previously for quantum-enhanced target detection \cite{lloyd08quantumillumination, tan08qigaussianstates, guha09quantumilluminationOPA, zhang15eesensing, shapiro20QIstory, nair20qi}, our transmitter generates $n$ TMSV states, and probes the target's transmittance with one mode of each TMSV state, while retaining the other mode as a reference.
In Section \ref{sec:optimal_receiver}, we characterize a matching quantum receiver, that measures the $n$ returned probes and corresponding entangled reference signals, and applies maximum likelihood estimation (MLE) on the resulting classical measurements.
In the limit of low transmitted power and large $n$ this transceiver achieves the fundamental lower bound on MSE from \cite{nair20qi}.
Although they are not classical, the components used in our receiver are well-known to the optics community: a two-mode squeezer followed by a photon-number-resolving (PNR) measurement.
However, despite this convenience, our receiver's existence is limited to certain ranges of system parameters: transmittance, signal power, and thermal noise power.
Thus, in Section \ref{sec:other_sensors}, we compare its theoretical limits to those of other well-known receivers, and show significant advantage derived from using TMSV input and our receiver.

The MSE of our sensing system converges to the quantum lower bound as the number of probes $n\to\infty$.
However, practical sensing is limited to a finite number of probes: $n<\infty$.
This motivates evaluating the speed of convergence to the limit.
Further complicating the analysis is the dependence of our receiver's structure on the parameter of interest, transmittance.
This is common in quantum estimation, and is addressed using the two-stage method from \cite[Ch.~6.4]{hayashi17qit}, \cite{gill00twostagemeasurement,hayashi05twostagemeasurement}.
Unfortunately, this complicates the analysis.
In Section \ref{sec:simulations} we use Monte Carlo simulation to study the performance of transmittance sensing using our sensor, and two other receivers, when the number of probes is finite. 
We show that the MSEs for these systems converge rapidly to their corresponding quantum limits, even when the two-stage method is used.

The transceiver derived here is optimal in the low-transmitted-power regime.
This corresponds to previous QI results, which demonstrated quantum advantage in this regime \cite{lloyd08quantumillumination, tan08qigaussianstates, guha09quantumilluminationOPA, sanz17estimationQI, shapiro20QIstory, nair20qi}.
Nevertheless, this has significant practical applications to sensors operating under total transmitted power constraints.
These can be imposed by, e.g., covertness/low probability of detection (LPD) \cite{bash12sqrtlawisit, bash13squarerootjsacnonote, bash15covertcommmag, bash15covertbosoniccomm, bash17qcovertsensingisit, gagatsos19floodlightsensor, bullock20discretemod, gagatsos20codingcovcomm}, battery size, or the need to limit exposure of a biological sample to light.
Although technically challenging, an experimental validation of our design is feasible in the near term, as the required squeezing and PNR measurement have been demonstrated.
We conclude with a discussion of future work in Section \ref{sec:discussion}.

\emph{Note:} The fundamental limits of transmittance estimation using TMSV in Section \ref{sec:tmsv} and the preliminary comparison to other systems in Section \ref{sec:other_sensors} were presented at the 2021 International Symposium on Information Theory (ISIT 2021) and published in its proceedings \cite{gong21losssensing}.  Receiver design and simulations in Sections \ref{sec:optimal_receiver} and \ref{sec:simulations} are new, while Section \ref{sec:other_sensors} is significantly expanded from \cite{gong21losssensing}.

\section{Prerequisites}
\label{sec:prerequisites}

\subsection{Notation}
As is customary in quantum information theory, we indicate operators with hats, e.g., $\hat{a}$ and $\hat{\rho}$.
Conjugate transpose and trace of $\hat{a}$ are indicated by $\hat{a}^\dag$ and $\trace\left\{\hat{a}\right\}$.
Density operators, which describe quantum states, are denoted using Greek letters, e.g., $\hat{\rho}$ and $\hat{\sigma}$.
We employ ket $\ket{x}$ and bra $\bra{x}\equiv\ket{x}^\dag$ notation for pure quantum states.
A particularly useful pure state is a Fock, or photon number, state $\ket{n}$ which represents exactly $n$ photons.
We employ caret to indicate classical estimators that input measured data and output an estimate, e.g., $\check{\theta}$ is an estimator for $\theta$.
We denote mean quantities with an overbar, e.g., $\bar{n}$ often indicates mean photon number.

\subsection{Channel Model}
\label{sec:channel}
A classical linear channel relates the complex-valued input and output amplitude mode functions $a$ and $b$ by $b=\sqrt{\theta}a+z$, where $\theta$ is power transmittance and $z$ is additive noise. 
Using an independent Gaussian random variable to represent noise, $z$, yields an additive white Gaussian noise (AWGN) channel, and simplifies estimation of transmittance $\theta$ \cite{kay93statSP1, vantrees01part1}.
However, optical, microwave, and radio-frequency (RF) light used for communications and sensing is fundamentally an electromagnetic wave described quantum mechanically by a boson field.
Noises of quantum-mechanical origin limit the performance of advanced high-sensitivity photodetection systems \cite{sinclair19nanowire,mccaughan19nanowire,lee18tes}.
Therefore, achieving the ultimate limits of estimation generally requires tools from quantum-information processing \cite{helstrom76quantumdetect, nielsen00quantum, wilde16quantumit2ed, hayashi17qit} and quantum optics \cite{scully97quantumoptics,agarwal12quantumoptics,
 orszag16quantumotpics}.
 
Formally, a single-mode lossy thermal-noise bosonic channel $\mathcal{E}^{\bar{n}_{\rm T},\theta}_{S\rightarrow R}$ in Fig.~\ref{fig:setup} describes quantum mechanically the transmission of a single (spatio-temporal-polarization) mode of
the electromagnetic field at a given transmission wavelength (such as optical, microwave, RF) over linear loss and additive Gaussian noise (such as noise stemming from blackbody radiation).
A beamsplitter with transmittance $\theta$ models power loss.
In contrast to the classical linear model, here the input-output relationship $\hat{b}=\sqrt{\theta}\hat{a}+\sqrt{1-\theta}\hat{e}$ between the bosonic modal annihilation operators of the single-mode channel requires the environment mode $\hat{e}$ to preserve the 
unitarity
by ensuring that commutator $\left[\hat{b},\hat{b}^\dag\right]=\hat{I}$, where $\hat{I}$ is the identity operator.
Excess noise is modeled by mode $\hat{e}$ being in a zero-mean thermal state $\hat{\tau}_{\bar{n}_{\rm T}}$, where $\bar{n}_{\rm T}$ is the mean photon number per mode injected by the environment.
Thermal state is represented in Fock (photon number) basis as $\hat{\tau}_{\bar{n}_{\rm T}}=\sum_{k=0}^{\infty}q\left(k;\bar{n}_{\rm T}\right)\ket{k}\bra{k}$, where the Bose-Einstein probability mass function
\begin{align}
\label{eq:tk}q(k;\bar{n})&=\frac{\bar{n}^k}{(1+\bar{n})^{k+1}}
\end{align}
is a variant of a geometric distribution.
In this paper, we are interested in estimating unknown transmittance $\theta$.
Before we state our problem formally in Section \ref{sec:problemstatement}, we review the concepts from quantum estimation theory that we require.

\begin{figure}[h]
\centering
	\includegraphics[width=0.5\columnwidth]{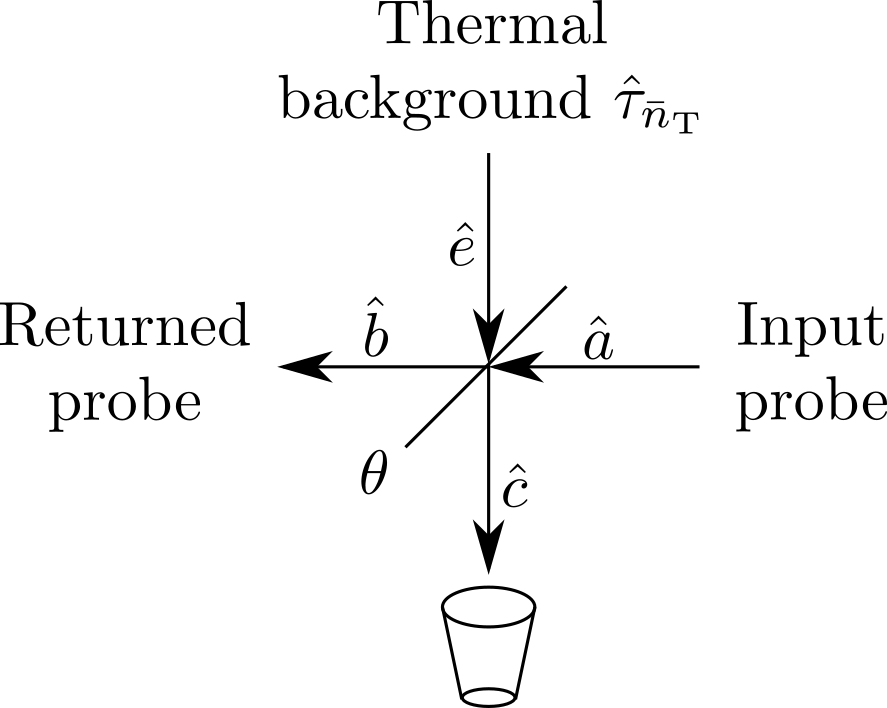}
	\caption{Single-mode thermal-noise lossy bosonic channel $\mathcal{E}^{(\bar{n}_{\rm T},\theta)}$ 
	modeled by a beamsplitter with transmittance $\theta$.  The sensor injects a probe in one of the input ports, while an environment 
	injects a thermal state $\hat{\tau}_{\bar{n}_{\rm T}}$ with mean photon
	number $\bar{n}_{\rm T}$ in the other input port.  The sensor has access to one of the output ports while the other output is absorbed by the environment. $\hat{a}$, $\hat{e}$, $\hat{b}$, and $\hat{c}$ 
	label the corresponding input/output modal annihilation operators.
	 \label{fig:bosonic}}
	\vspace{-0.1cm}
\end{figure}

\subsection{Introduction to Quantum Estimation} \label{sec:quantum_estimation}

Suppose a quantum state $\hat{\sigma}(\theta)$ physically encodes information about an unknown parameter $\theta$ (transmittance in this paper).
We are interested in estimating $\theta$ from $\hat{\sigma}(\theta)$.
A physical device that extracts information about $\theta$ from $\hat{\sigma}(\theta)$ is described by a positive operator-valued measure (POVM) $\left\{\hat{\Gamma}_x\right\}$ that satisfies the non-negativity and completeness conditions:
\begin{align}
    \forall x: \hat{\Gamma}_x\ge0, \ \ \sum_x \hat{\Gamma}_x = \hat{I},
\end{align}
where $\hat{I}$ is the identity operator.  For example, an ideal photon-number-resolving (PNR) measurement is described by $\left\{\ket{x}\bra{x}\right\}$, where $\ket{x}$ is the Fock (photon number) state.  
Classical statistics of an output of a device characterized by POVM $\left\{\hat{\Gamma}_x\right\}$ are described by a random variable $X$ with probability mass function $p_{X}(x;\theta) = \trace\left\{\hat{\Gamma}_x\hat{\sigma}(\theta)\right\}$ \cite[Ch.~III]{helstrom76quantumdetect}.

An estimator $\check{\theta}(x)$, is a function of the observation $x$, which is an instance of $X$.  We desire an unbiased $\check{\theta}(x)$, i.e., $E_X\left[\check{\theta}(X)\right]=\theta_0$, that minimizes the mean square error (MSE)  $V_{\theta_0}\left(\check{\theta}\right)=E_X\left[\left(\check{\theta}(X)-\theta_0\right)^2\right]$, where $\theta_0$ is the true value of $\theta$ and $E_X[f(X)]$ is the expected value of $f(X)$.  The classical Cram\'{e}r-Rao bound (CRB) lower bounds the MSE:
\begin{align}\label{eq:cCRB}V_{\theta_0}\left(\check{\theta}\right)&\ge\frac{1}{\mathcal{I}_{\theta}(X)},
\end{align}
where $ \mathcal{I}_{\theta}(X)=E_{X}\left[\left(\partial \log p_X(X;\theta)/\partial\theta\right)^2 \right] $ is the classical Fisher information (FI) associated with $\theta$ for random variable $X$. By additivity of classical FI, for a sequence of $n$ independent and identically distributed (i.i.d.) random variables $\left\{X_k\right\}_{k=1}^n$, $\mathcal{I}_{\theta}\left(\left\{X_k\right\}_{k=1}^n\right)=n\mathcal{I}_{\theta}\left(X_1\right)$. Estimators that achieve the classical CRB in \eqref{eq:cCRB} with equality are called \emph{efficient}. For observations described by an i.i.d.~sequence of random variables $\left\{X_k\right\}_{k=1}^n$, maximum likelihood estimator (MLE) is asymptotically unbiased and efficient as $n\to\infty$, up to mild regularity conditions \cite{kay93statSP1, vantrees01part1}.

However, classical estimation theory implicitly assumes that the  measurement device (described by POVM $\left\{\hat{\Gamma}_x\right\}$) is fixed. 
Quantum methodology enriches estimation theory by allowing analysis and optimization of $\left\{\hat{\Gamma}_x\right\}$ \cite[Ch.~VIII]{helstrom76quantumdetect}. The quantum CRB lower bounds the classical CRB since it assumes an optimal measurement: 
\begin{align}\label{eq:qCRB}V_{\theta_0}\left(\check{\theta}\right)&\ge\frac{1}{\mathcal{J}_{\theta}\left(\hat{\sigma}(\theta)\right)},
\end{align}
where $\mathcal{J}_{\theta}\left(\hat{\sigma}(\theta)\right) = \trace\left\{\hat{\Lambda}_{\theta}^2\hat{\sigma}(\theta)\right\}$ is the quantum FI associated with $\theta$ for state $\hat{\sigma}(\theta)$.  $\hat{\Lambda}_{\theta}$ is the symmetric logarithm derivative (SLD) operator that is Hermitian but not necessarily positive.  It is defined implicitly by \cite[Ch.~VIII.4(b)]{helstrom76quantumdetect}:
\begin{align}
\label{eq:sld_def}  \frac{\partial\hat{\sigma}(\theta)}{\partial{\theta}} = \frac{1}{2}\left(\hat{\Lambda}_{\theta}\hat{\sigma}(\theta)+\hat{\sigma}(\theta)\hat{\Lambda}_{\theta}\right).
\end{align}
Analogous to classical FI, by the additivity of quantum FI, for a tensor product of $n$ states $\hat{\sigma}^{\otimes n}(\theta)$, $\mathcal{J}_{\theta}\left(\hat{\sigma}^{\otimes n}(\theta)\right)=n\mathcal{J}_{\theta}\left(\hat{\sigma}(\theta)\right)$.
We call a combination of a quantum measurement and a classical estimator on the corresponding output \emph{quantum efficient}, if it achieves the quantum CRB in \eqref{eq:qCRB} with equality.

Consider an eigendecomposition of SLD $\hat{\Lambda}_{\theta}=\sum_x\lambda_x(\theta)\ket{\lambda_x(\theta)}\bra{\lambda_x(\theta)}$, where $\left\{\ket{\lambda_x(\theta)}\right\}$ is a set of orthonormal pure eigen-states of $\hat{\Lambda}_{\theta}$ and $\left\{\lambda_x(\theta)\right\}$ are the corresponding eigenvalues.
For a tensor-product state $\hat{\sigma}^{\otimes n}(\theta)$, a tensor-product measurement $\left\{\ket{\lambda_x(\theta)}\bra{\lambda_x(\theta)}\right\}^{\otimes n}$ constructed from eigendecomposition of SLD $\hat{\Lambda}_{\theta}$ and followed by an MLE on the corresponding classical i.i.d.~random output sequence $\left\{X_k\right\}_{k=1}^n$ is asymptotically unbiased and quantum efficient.  Thus, unlike in optimal decoders for general classical-quantum channels \cite{Hol97}, a complicated joint-detection measurement that entangles the output of $n$ probes is unnecessary: $\left\{\ket{\lambda_x(\theta)}\bra{\lambda_x(\theta)}\right\}$ is applied separately to each of $n$ states $\hat{\sigma}(\theta)$.
However, it is important to note that there is an infinite number of eigendecompositions of $\hat{\Lambda}_{\theta}$, and that their mathematical expressions are typically unavailable in closed form.  Even when they are found, translating these expressions to physical devices can be extremely challenging.

Furthermore, the structure of a quantum CRB-achieving measurement $\left\{\ket{\lambda_x(\theta)}\bra{\lambda_x(\theta)}\right\}$ often depends on the parameter of interest $\theta$.
This seeming paradox of needing to know $\theta$ to build a device for its measurement is addressed by the following two-stage approach \cite[Ch.~6.4]{hayashi17qit}, \cite{gill00twostagemeasurement,hayashi05twostagemeasurement}. First, one obtains a rough pre-estimate $\check{\theta}_0$ using $n^\beta$, $\beta\in\left(\frac{1}{2},1\right)$, probes and a sub-optimal measurement that does not depend in $\theta$.
Then, one employs $\check{\theta}_0$ to construct $\left\{\ket{\lambda_x(\check{\theta}_0)}\bra{\lambda_x(\check{\theta}_0)}\right\}$ and refine $\check{\theta}$ using the remaining $n-n^\beta$ probes.
This achieves quantum CRB asymptotically as $n\to\infty$, under conditions outlined in \cite[Ch.~6.4]{hayashi17qit}, \cite{hayashi05twostagemeasurement}.

We encourage the reader, interested in learning more about the foundations of quantum sensing, to consult the classic texts on quantum detection and estimation \cite{helstrom76quantumdetect} and quantum information theory \cite{nielsen00quantum}, as well as more recent texts \cite{wilde16quantumit2ed, hayashi17qit} covering these subjects in greater depth.
We are now ready for the formal treatment of quantum transmittance estimation.

\begin{figure}[h]
\centering
	\includegraphics[width=0.8\columnwidth]{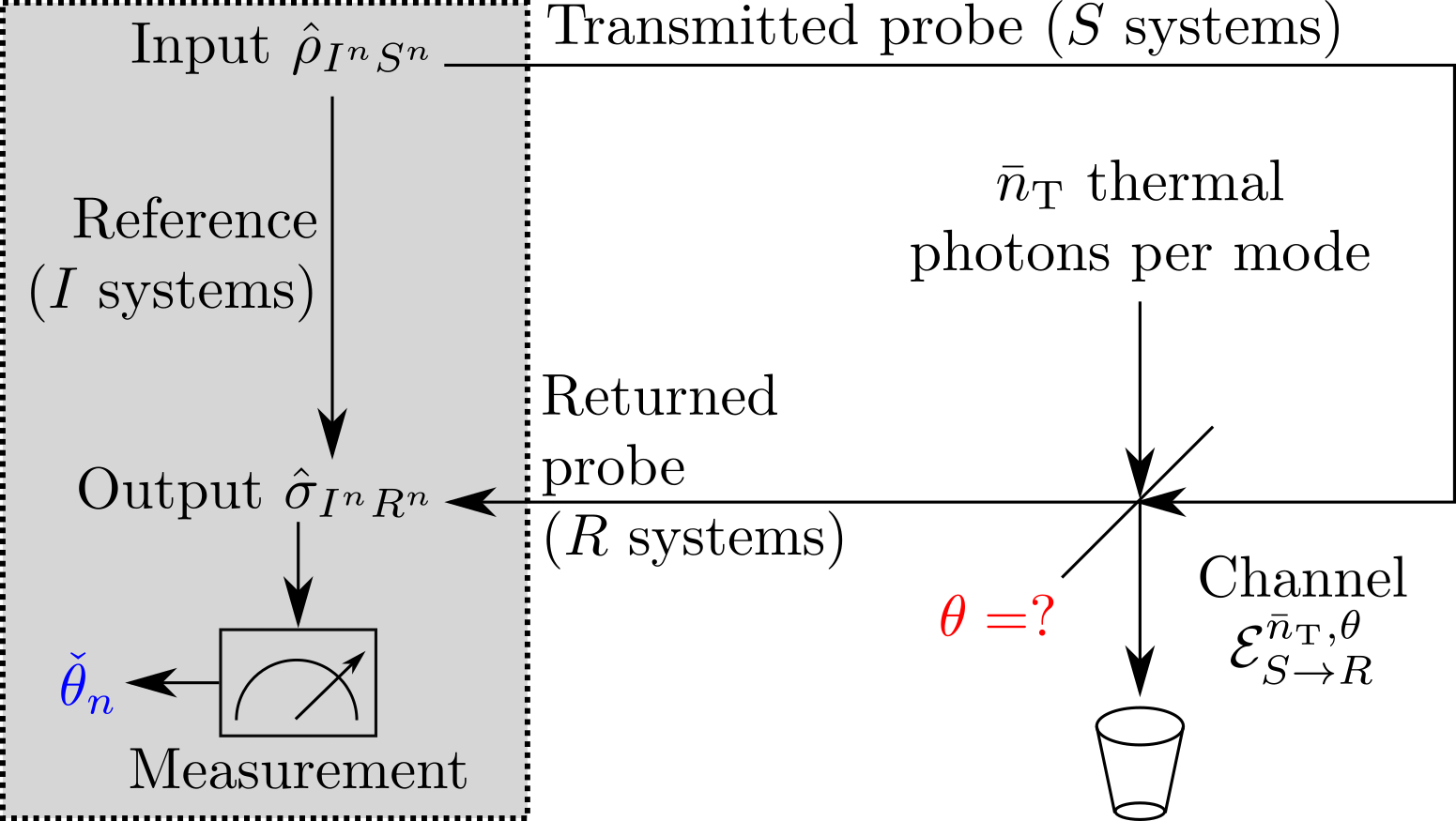}
	\caption{Sensing of unknown transmittance $\theta$. Sensor transmits $n$-mode signal systems $S$ of bipartite state $\hat{\rho}_{I^nS^n}$ by using $n$ times the single-mode thermal noise lossy bosonic channel $\mathcal{E}^{(\bar{n}_{\rm T},\theta)}$ 
	described in Fig.~\ref{fig:bosonic}, where mean thermal photon
	number $\bar{n}_{\rm T}\equiv\frac{\bar{n}_{\rm B}}{1-\theta}$. 	Idler systems $I$ are retained as a reference for the measurement of output state $\hat{\sigma}_{I^nR^n}(\theta)$, which yields the estimator $\check{\theta}_n$.
	 \label{fig:setup}}
	\vspace{-0.1cm}
\end{figure}

\subsection{Quantum Transmittance Estimation}
\label{sec:problemstatement}

Fig.~\ref{fig:setup} depicts our system setup.
Sensor's goal is to estimate unknown power transmittance $\theta$.
It prepares a bipartite quantum state $\hat{\rho}_{I^nS^n}$ which occupies $n$ signal systems $S$ and $n$ idler systems $I$.
Signal systems $S$ interrogate the target over $n$ available modes of channel $\mathcal{E}^{\bar{n}_{\rm T},\theta}_{S\rightarrow R}$.
Idler systems $I$ are retained losslessly and noiselessly as a reference.
The output state $\hat{\sigma}_{I^nR^n}(\theta)=\left(\mathcal{I}_{I}\otimes\mathcal{E}^{\bar{n}_{\rm T},\theta}_{S\rightarrow R}\right)^{\otimes n}\left[\hat{\rho}_{I^nS^n}\right]$ carries information about transmittance $\theta$ in the returned systems $R$, where $\mathcal{I}$ is the identity channel.
As described in Section \ref{sec:quantum_estimation}, we seek an unbiased estimator $\check{\theta}_n$ on a measurement of $\hat{\sigma}_{I^nR^n}(\theta)$ that minimizes the mean squared error $V_{\theta_0}\left(\check{\theta}_n\right)$.
Early work focused on the transmittance sensing in a pure-loss bosonic channel $\mathcal{E}^{0,\theta}_{S\rightarrow R}$, with $\bar{n}_{\rm T}=0$ \cite{monras07loss, adesso09optlossest, nair18loss}.
Notably, the author of \cite{nair18loss} proved that Fock (photon number) states are optimal for transmittance sensing in this environment.

Unfortunately, the pure-loss bosonic channel model has limited applications, due to the omnipresence of thermal noise in practical scenarios. 
Quantum illumination (QI) improves detection and estimation in thermal noise using entanglement between $S$ and $I$.
The quantum CRB for joint estimation of unknown $\theta$ and $\bar{n}_{\rm T}$ using Gaussian subset of quantum states  \cite{weedbrook12gaussianQIrmp} was derived in \cite{monras11tmsvopt}.
Although the two-mode squeezed vacuum (TMSV) state was proved an optimal Gaussian state in \cite{monras11tmsvopt}, the sensor was allowed to estimate $\theta$ from just the thermal background -- called ``shadow effect'' in \cite{jonsson22gaussianlosssensing}. 
QI literature \cite{nair20qi, lloyd08quantumillumination, tan08qigaussianstates, guha09quantumilluminationOPA, shapiro20QIstory, sanz17estimationQI} addresses arguably more practical settings where the thermal background cannot aid the estimation. 
The ``shadow effect'' is removed by setting: 
\begin{align}
\label{eq:barnT}\bar{n}_{\rm T}&\equiv\frac{\bar{n}_{\rm B}}{1-\theta},
\end{align}
where $\bar{n}_{\rm B}$ is the mean number of thermal background photons per mode that corrupt the sensor's probes.
When the background light does not help the estimation of $\theta$, quantum CRB is quantitatively different from results in \cite{monras11tmsvopt}\footnote{Reparameterizing \cite[Eqs.~(B8a)-(B8d)]{monras11tmsvopt} to $\theta=e^{-\gamma}$ shows that, even in the absence of probes ($\bar{n}_{\rm S}=0$), quantum FI associated with $\theta$ is positive due to the thermal background.}, and the quantum FI is upper-bounded by \cite[Eq.~(21)]{nair20qi}:
\begin{align}
\label{eq:ubQFI}\mathcal{J}_{\theta}\left(\hat{\sigma}_{I^nR^n}(\theta)\right)&\leq\mathcal{J}_{\theta,\mathrm{ub}}\left(\bar{n}_{\rm S,tot}\right)\equiv\frac{\bar{n}_{\rm S,tot}}{\theta(\bar{n}_{\rm B}+1-\theta)},
\end{align}
where $\bar{n}_{\rm S,tot}=\sum_{i=1}^n\bar{n}_{\mathrm{S},i}$ is the total mean photon number transmitted over $n$ modes, and the individual mode mean photon numbers $\bar{n}_{\mathrm{S},i}$ may be unequal.
However, \cite{nair20qi} leaves open the structure of $\hat{\rho}_{I^nS^n}$ that saturates \eqref{eq:ubQFI}, and the design of the corresponding quantum-CRB-achieving measurement.

Excited by the gap identified in \cite{nair20qi} and using the prescription for $\bar{n}_{\rm T}$ in \eqref{eq:barnT}, we analyze transmittance sensing with the TMSV states.
In the next section, we find that the quantum FI of TMSV states saturates the ultimate bound in \eqref{eq:ubQFI} as per-mode transmitted photon number $\bar{n}_{\rm S}\to 0$.
In Section \ref{sec:other_sensors}, we report that, at $\bar{n}_{\rm S}>0$, the quantum FI of TMSV states significantly exceeds the FI of other well-known transmittance estimation schemes.  
This motivates the derivation of the quantum-CRB-achieving receiver structure for TMSV probes in Section \ref{sec:optimal_receiver}, and the numerical analysis of the convergence of its MSE to the optimal in Section \ref{sec:simulations}.

Following the conference presentation \cite{gong21losssensing} of our preliminary results, \cite{jonsson22gaussianlosssensing} investigated transmittance sensing with and without noise aiding the estimation.
Notably, \cite{jonsson22gaussianlosssensing} proved that TMSV maximized the quantum FI within the class of Gaussian quantum states \cite{weedbrook12gaussianQIrmp} with and without the ``shadow effect.''
While the treatment of quantum FI in \cite{jonsson22gaussianlosssensing} is comprehensive, the receiver structures that achieve it are not considered.
Even more recently, \cite{goldberg22multiparameterlossxanadu} presented an experimental study of estimating multiple transmittance parameters in Xanadu's X8 integrated-photonic-quantum computer \cite{arrazola21photonicqc}.
Unfortunately, the limitations of Xanadu's platform limit the study in \cite{goldberg22multiparameterlossxanadu} to photon-number-resolving (PNR) measurements of TMSV states.
The same authors follow up by analyzing the limits of transmittance sensing using coherent and Fock states in the presence of detector dark counts \cite{goldberg22lossdarkcounts}, and report results that are qualitatively similar to those in Section \ref{sec:other_sensors}.

\section{TMSV is Optimal for Transmittance Sensing}
\label{sec:tmsv}

The TMSV state is represented in the Fock (photon number) basis as follows:
\begin{align}
\label{eq:tmsv}\ket{\psi}_{IS}&=\sum_{k=0}^\infty\sqrt{q(k;\bar{n}_{\rm S})}\ket{k}_{I}\ket{k}_{S},
\end{align}
where $q\left(k;\bar{n}\right)$ is defined in \eqref{eq:tk}.
TMSV is a zero-displacement pure Gaussian state, which among all two-mode-Gaussian states with mean photon number $\bar{n}_{\rm S}$  is maximally entangled \cite{weedbrook12gaussianQIrmp}.
It is critical in quantum-information processing.
Generating TMSV is a well-known (bordering on routine) process in quantum optics. 
We show that TMSV becomes optimal for transmittance estimation in thermal noise as transmitted photon number per mode $\bar{n}_{\rm S}$ decays to zero:

\begin{theorem}
\label{th:tmsv} 
The following limit holds for the quantum Fisher information $\mathcal{J}_{\theta}\left(\hat{\sigma}_{I^nR^n}(\theta)\right)$:
\begin{align}
\label{eq:limit_tmsv}\lim_{\bar{n}_{\rm S}\rightarrow0}\frac{\mathcal{J}_{\theta}\left(\hat{\sigma}_{I^nR^n}(\theta)\right)}{\bar{n}_{\rm S}}&=\frac{n}{\theta(\bar{n}_{\rm B}+1-\theta)}
\end{align} 
when $n$ TMSV probes described by tensor-product state $\hat{\rho}_{I^nS^n}=\ket{\psi}\bra{\psi}_{IS}^{\otimes n}$ are used and $\hat{\sigma}_{I^nR^n}(\theta)$ is the quantum state describing the returned probes and retained references.
\end{theorem}

We first note that the lossy thermal-noise bosonic channel $\mathcal{E}^{\bar{n}_{\rm T},\theta}_{S\rightarrow R}$ acts independently on each transmitted mode.
Therefore, for input tensor product of TMSV states $\ket{\psi}\bra{\psi}_{IS}^{\otimes n}$, the output state $\hat{\sigma}_{I^nR^n}(\theta)=\hat{\sigma}_{IR}^{\otimes n}(\theta)$ is a tensor product of states $\hat{\sigma}_{IR}(\theta)=\left(\mathcal{I}_{I}\otimes\mathcal{E}^{\bar{n}_{\rm T},\theta}_{S\rightarrow R}\right)\left[\ket{\psi}\bra{\psi}_{IS}\right]$.
By the additivity of the quantum FI for tensor product states, $\mathcal{J}_{\theta}\left(\hat{\sigma}_{I^nR^n}(\theta)\right)=n\mathcal{J}_{\theta}\left(\hat{\sigma}_{IR}(\theta)\right)$.
In Appendix \ref{app:tmsv} we employ the method from \cite{Marian2012,Banchi2015} to derive the quantum FI 
\begin{IEEEeqnarray}{rCl}
\IEEEeqnarraymulticol{3}{l}{\mathcal{J}_{\theta,\text{TMSV}}\left(\bar{n}_{\rm S}\right)\equiv\mathcal{J}_{\theta}\left(\hat{\sigma}_{IR}(\theta)\right)}\IEEEnonumber\\
\label{eq:tmsv_qfi}&=&\frac{\bar{n}_{\rm S}\left(\bar{n}_{\rm B}+1+(1-\theta)\bar{n}_{\rm S}+\bar{n}_{\rm B}\bar{n}_{\rm S}\right)}{\theta\left(\bar{n}_{\rm B}+1-\theta\right)\left(\bar{n}_{\rm B}+1+\bar{n}_{\rm S}(2\bar{n}_{\rm B}+1-\theta)\right)}
\end{IEEEeqnarray}
associated with the quantum state $\hat{\sigma}_{IR}(\theta)$ that describes the returned probe and retained reference when a single TMSV probe is used.
Multiplying \eqref{eq:tmsv_qfi} by $\frac{n}{\bar{n}_{\rm S}}$ and taking the limit in \eqref{eq:limit_tmsv} yields the proof.\footnote{Our expression for the quantum FI of TMSV in \eqref{eq:tmsv_qfi} is exact, unlike \cite[Eq.~(6)]{sanz17estimationQI} and \cite[Eq.~(23)]{nair20qi}. In fact, all of the quantum FI expressions in \cite{sanz17estimationQI} are approximations at $\theta_{\rm a}=0$, where $\theta_{\rm a}=\sqrt{\theta}$ is the amplitude transmittance, and \cite[Eq.~(23)]{nair20qi} is \cite[Eq.~(6)]{sanz17estimationQI} reparametrized from $\theta_{\rm a}$ to $\theta$.
Reparametrizing \eqref{eq:tmsv_qfi} to $\theta_{\rm a}$ and setting $\theta_{\rm a}=0$ yields \cite[Eq.~(6)]{sanz17estimationQI}.
That a crude approximation with a zero-order Taylor series term yields such a close result is striking.}
  
Theorem \ref{th:tmsv} proves that TMSV is optimal over \emph{all} low-input-power states, including the non-Gaussian ones.  
Although it has been shown to be an optimal Gaussian state for all values of $\theta$, $\bar{n}_{\rm B}$, and $\bar{n}_{\rm S}$ \cite{jonsson22gaussianlosssensing}, characterization of a general quantum input state that maximizes quantum FI associated with $\theta$ is an open problem.
That being said, as mentioned in Section \ref{sec:intro}, the low transmitted-photon-number per mode regime is important for the design of sensors operating under the total power constraints.
Thus, we characterize and analyze a receiver structure that achieves asymptotically the quantum CRB for TMSV.

\section{Transmittance Estimation Using TMSV Probes}
\label{sec:optimal_receiver}

\subsection{Optimal Receiver Structure}
We determine the eigendecomposition of the symmetric logarithmic derivative (SLD) operator $\hat{\Lambda}_{\theta}$ defined in \eqref{eq:sld_def} for the quantum state $\hat{\sigma}_{IR}(\theta)$  describing the returned probe and retained reference.
Since $\hat{\sigma}_{IR}(\theta)$ is Gaussian, $\hat{\Lambda}_{\theta}$ is a degree-2 polynomial of creation and annihilation operators $\hat{a}_{I}$, $\hat{a}_{R}$, $\hat{a}_{I}^\dag$, and $\hat{a}_{R}^\dag$ \cite[Sec.~III]{jiang14qfigaussian}.
Thus, there are infinitely-many unitary transformations $\hat{U}_{\theta}$ that represent the application of a finite sequence of squeezing, displacement, phase rotation, and beam splitter operators, that diagonalize $\hat{\Lambda}_{\theta}$.  

\begin{figure}[h]
\centering
	\includegraphics[width=0.8\columnwidth]{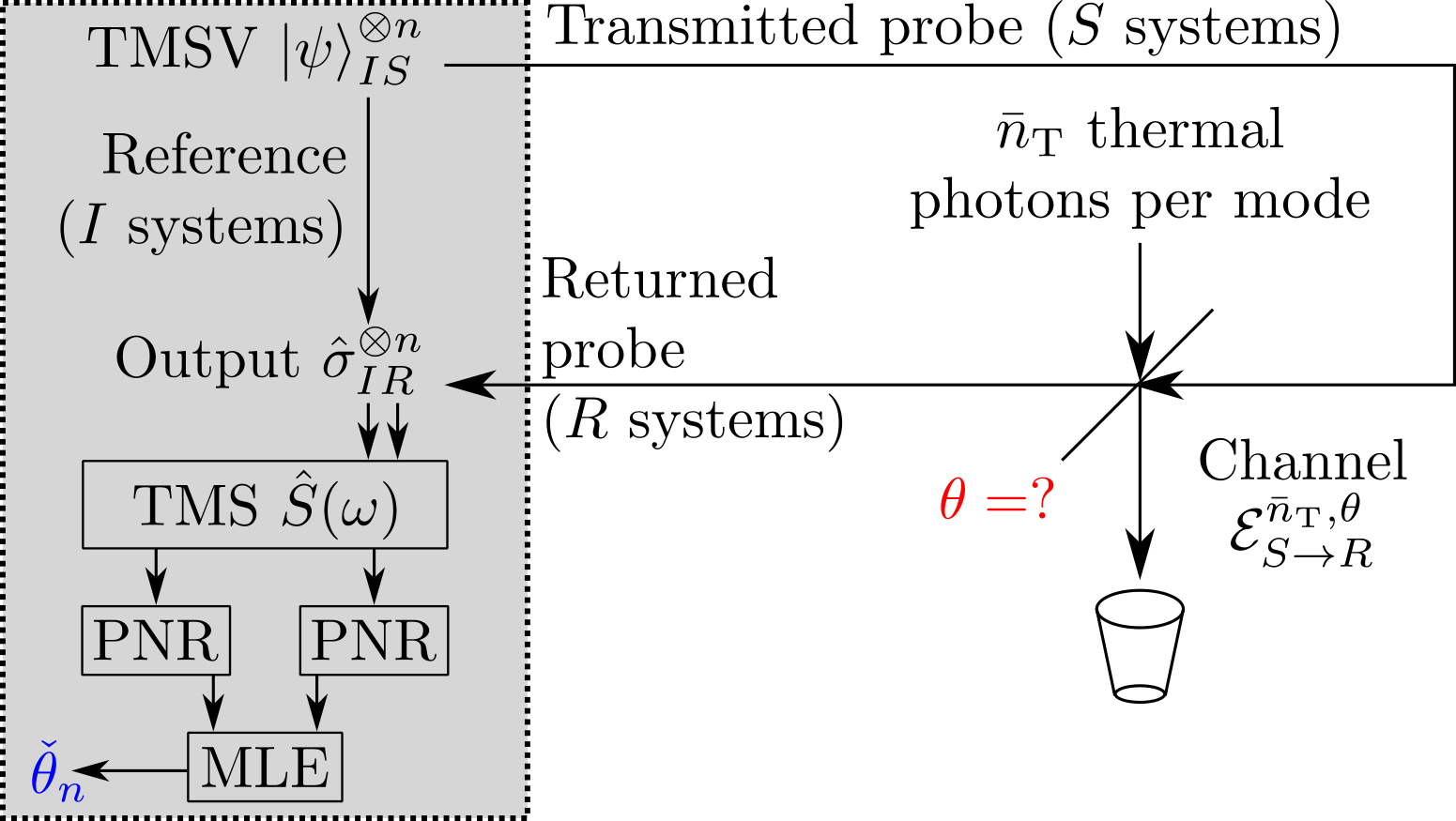}
	\caption{Sensing the unknown transmittance $\theta$ using $n$ two-mode squeezed vacuum (TMSV) states $\ket{\psi}_{IS}$.  When $
\ket{\psi}_{IS}$ is transmitted, a bipartite output state $\hat{\sigma}_{IR}$ occupies a retained idler system $I$ and a corresponding returned probe system $R$.  The receiver achieves the quantum CRB by applying a two-mode squeezer (TMS) separately to each of the $n$ output states $\hat{\sigma}_{IR}$, followed by independent photon-number-resolving (PNR) measurement of each output mode.  Maximum likelihood estimator (MLE) is applied to the resulting classical output to obtain the estimate $\check{\theta}_n$. 
As in Fig.~\ref{fig:setup}, $\bar{n}_{\rm T}\equiv\frac{\bar{n}_{\rm B}}{1-\theta}$.
 	 \label{fig:optimal_receiver}}
	\vspace{-0.1cm}
\end{figure}

In Appendix \ref{app:optimal_receiver}, we adapt the approach from \cite{monras07loss} to show that $\hat{\Lambda}_{\theta}=c_1\hat{S}(\omega)\left(\hat{a}^{\dagger}_{I}\hat{a}_{I}+\hat{a}^{\dagger}_{R}\hat{a}_{R}\right)\hat{S}^\dag(\omega)+c_0\hat{I}$, where $\hat{S}(\omega)=\exp(\omega^\ast\hat{a}_R\hat{a}_I-\omega\hat{a}_R^\dagger\hat{a}_I^\dagger)$ is a two-mode squeezing operator, $\hat{I}$ is the identity operator, and $c_0$, $c_1$ are scalars.
Unitary $\hat{S}(\omega)$ diagonalizes $\hat{\Lambda}_{\theta}$ in the two-mode Fock (photon number) basis $\left\{\ket{km}\right\}\equiv\left\{\ket{k}\otimes\ket{m}\right\}$, $k,m=0,1,\ldots$ since this is an eigenbasis for $\hat{a}^{\dagger}_{I}\hat{a}_{I}+\hat{a}^{\dagger}_{R}\hat{a}_{R}$. Thus,  $\left\{\hat{S}(\omega)\ket{km}: k,m=0,1,\ldots\right\}$ is an eigendecomposition of $\hat{\Lambda}_{\theta}$.
Hence, a receiver for TMSV transmitter that achieves the quantum Cram\'{e}r-Rao bound (CRB) asymptotically is a two-mode squeezer followed by the photon-number-resolving (PNR) measurements of each output mode.
Fig.~\ref{fig:optimal_receiver} depicts transmittance sensing using TMSV and this receiver.
Formally, it is the positive operator-valued measurement (POVM) $\mathcal{M}=\left\{\ket{s_{km}}\bra{s_{km}}\right\}$, where $\ket{s_{km}}=\hat{S}(\omega)\ket{km}$.
POVM $\mathcal{M}$ followed by the maximum likelihood estimation (MLE) of $\theta$ from the classical outcome of PNR measurements achieves the quantum CRB as $n\to\infty$.
The squeezing parameter is $\omega=\lambda - \zeta$, where $\zeta$ and $\lambda$ are functions of $\theta$, $\bar{n}_{\rm B}$, and $\bar{n}_{\rm S}$ defined in \eqref{eq:zeta} and \eqref{eq:lambda}, respectively.
We emphasize that, although our measurement is convenient, in that it can be physically implemented using well-known optical elements (squeezer and PNR receivers), it is only \emph{one of the infinitely-many} measurements that achieve the quantum CRB asymptotically.

\subsection{Remarks and Caveats}
\label{sec:remarks_optimal_receiver}

The structure of our receiver is remarkable in that it achieves the quantum CRB using well-known optical components:
\begin{itemize}
\item Optical elements implementing a two-mode squeezer required for the receiver have been demonstrated in the laboratory \cite{eberle13tmsv10}.
\item Our Monte Carlo simulation described in Section \ref{sec:simulations} shows that, at a moderate amount of thermal noise ($\bar{n}_{\rm B}\approx 1$ photons per mode), and with a low transmitted photon number ($\bar{n}_{\rm S}\approx 0.01$), nine photons is the sufficient resolution for PNR measurements.  Such measurements, while technically complex, have been demonstrated \cite{miller03tes}.
\end{itemize}
However, two caveats are in order:

\subsubsection{Value of $\omega$ depends on $\theta$} As mentioned in Section \ref{sec:quantum_estimation}, the dependence of the structure of a quantum CRB-achieving measurement on the parameter of interest is a common occurrence in quantum estimation theory. 
In our Monte Carlo simulation detailed in Section \ref{sec:simulations} we employ the two-stage approach \cite[Ch.~6.4]{hayashi17qit},  \cite{gill00twostagemeasurement,hayashi05twostagemeasurement} described in Section \ref{sec:quantum_estimation} with coherent transceiver from Section \ref{sec:coherent_transceiver} used to construct a rough pre-estimate, $\check{\theta}_0$.

\begin{figure}[h]
\centering
	\includegraphics[width=1\columnwidth]{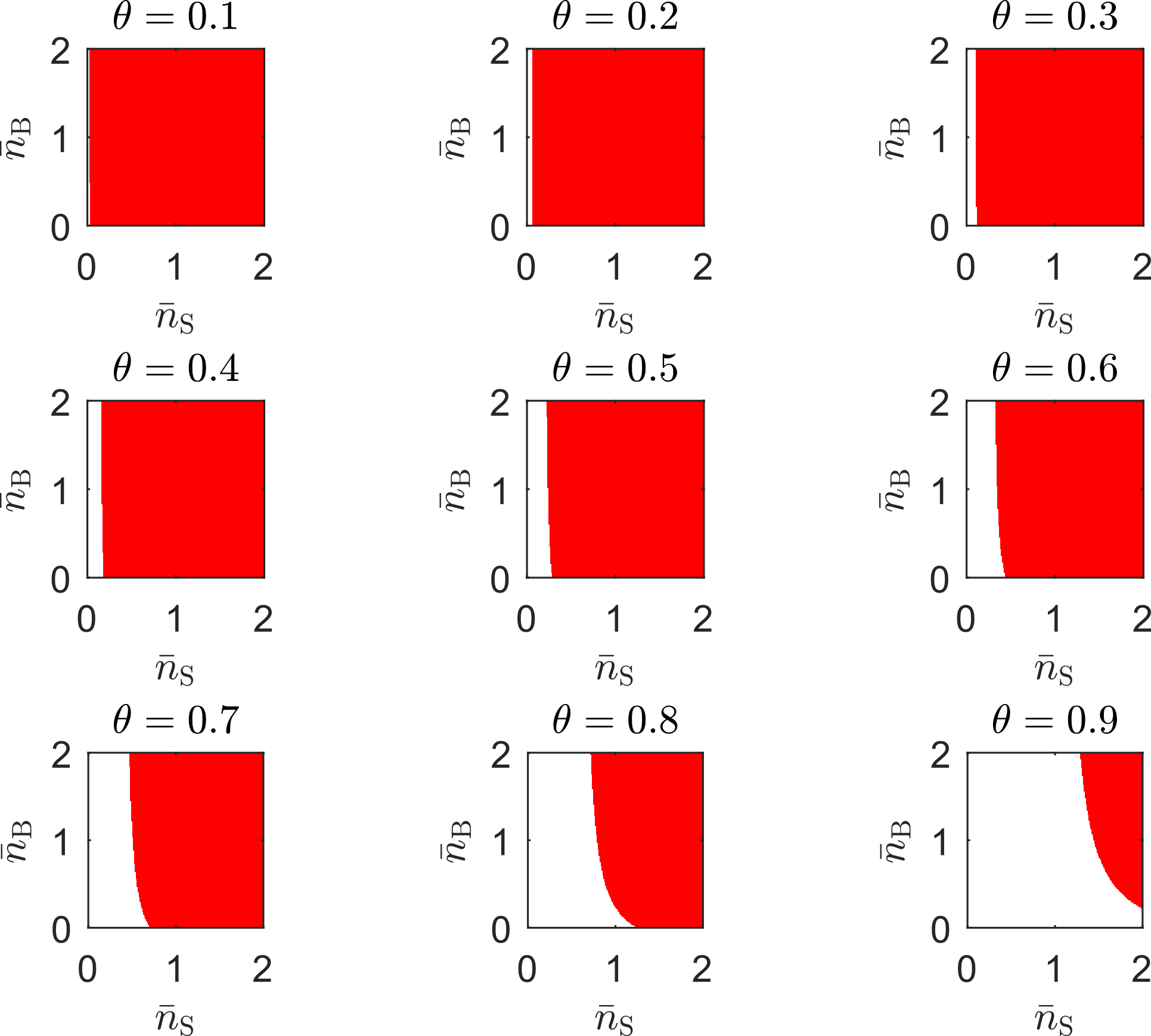}
	\caption{Satisfaction of the condition in \eqref{eq:receivercondition} for the existence of the receiver employing two-mode squeezing, followed by photon counting at various values of system parameters $\theta$, $\bar{n}_{\rm S}$, and $\bar{n}_{\rm B}$.  Red shading denotes the regions where \eqref{eq:receivercondition} is not satisfied and white the regions where \eqref{eq:receivercondition} is satisfied.
	 \label{fig:receivercondition}}
	\vspace{-0.1cm}
\end{figure}
\subsubsection{Existence of $\omega$ depends on $\theta$, $\bar{n}_{\rm S}$, and $\bar{n}_{\rm B}$} The eigenbasis of the SLD $\hat{\Lambda}_{\theta}$ in the form $\left\{\hat{S}(\omega)\ket{km}: k,m=0,1,\ldots\right\}$ requires the scalars $\lambda$, $F^{\prime}$, $T_1$, $T_2$ to be real (see Appendix \ref{app:eigenbasis}).  Thus, per \eqref{eq:tau1}-\eqref{eq:Fprime}, we must have: 
\begin{align}
\label{eq:receivercondition}(C+D)^2< 4 E^2,
\end{align}
where $C$, $D$, and $E$ are defined in \eqref{eq:C}-\eqref{eq:E}.
As Fig.~\ref{fig:receivercondition} illustrates,  \eqref{eq:receivercondition} does not hold for a certain range of system parameters: $\theta$, $\bar{n}_{\rm S}$, and $\bar{n}_{\rm B}$.
Therefore, another, likely more complicated, receiver structure is necessary to realize the full quantum advantage that the TMSV states yield.\footnote{In Appendix \ref{app:optimal_receiver} we adapt the approach in \cite{monras07loss}, where the authors derive a quantum CRB-achieving receiver for transmittance sensing when $\bar{n}_{\rm T}=0$.
That receiver similarly does not exist for certain ranges of $\bar{n}_{\rm S}$ and $\theta$.}

Notwithstanding these caveats, the range of operating parameters where our proposed receiver exists, corresponds to the low transmitted photon number regime ($\bar{n}_{\rm S}\ll 1$) which has significant practical applications, (e.g., in covert/low probability of detection (LPD) \cite{bash12sqrtlawisit, bash13squarerootjsacnonote, bash15covertcommmag, bash15covertbosoniccomm, bash17qcovertsensingisit, gagatsos19floodlightsensor, bullock20discretemod, gagatsos20codingcovcomm}, battery-constrained, or light-sensitive-sample scenarios).
It is also the regime where TMSV states are quantum optimal.
Nevertheless, we analyze and compare alternative transmittance-sensing approaches next.

\section{Comparison with Alternative Transmittance-Sensing Methods: Fundamental Limits}
\label{sec:other_sensors}

We compare the classical and quantum Fisher information (FI) associated with transmittance $\theta$ for several well-known receivers with the quantum FI for TMSV state derived in Section \ref{sec:tmsv} and the ultimate upper bound \eqref{eq:ubQFI} from \cite{nair20qi}.

\subsection{Coherent Transceiver}
\label{sec:coherent_transceiver}
\begin{figure}[h]
\centering
	\includegraphics[width=0.8\columnwidth]{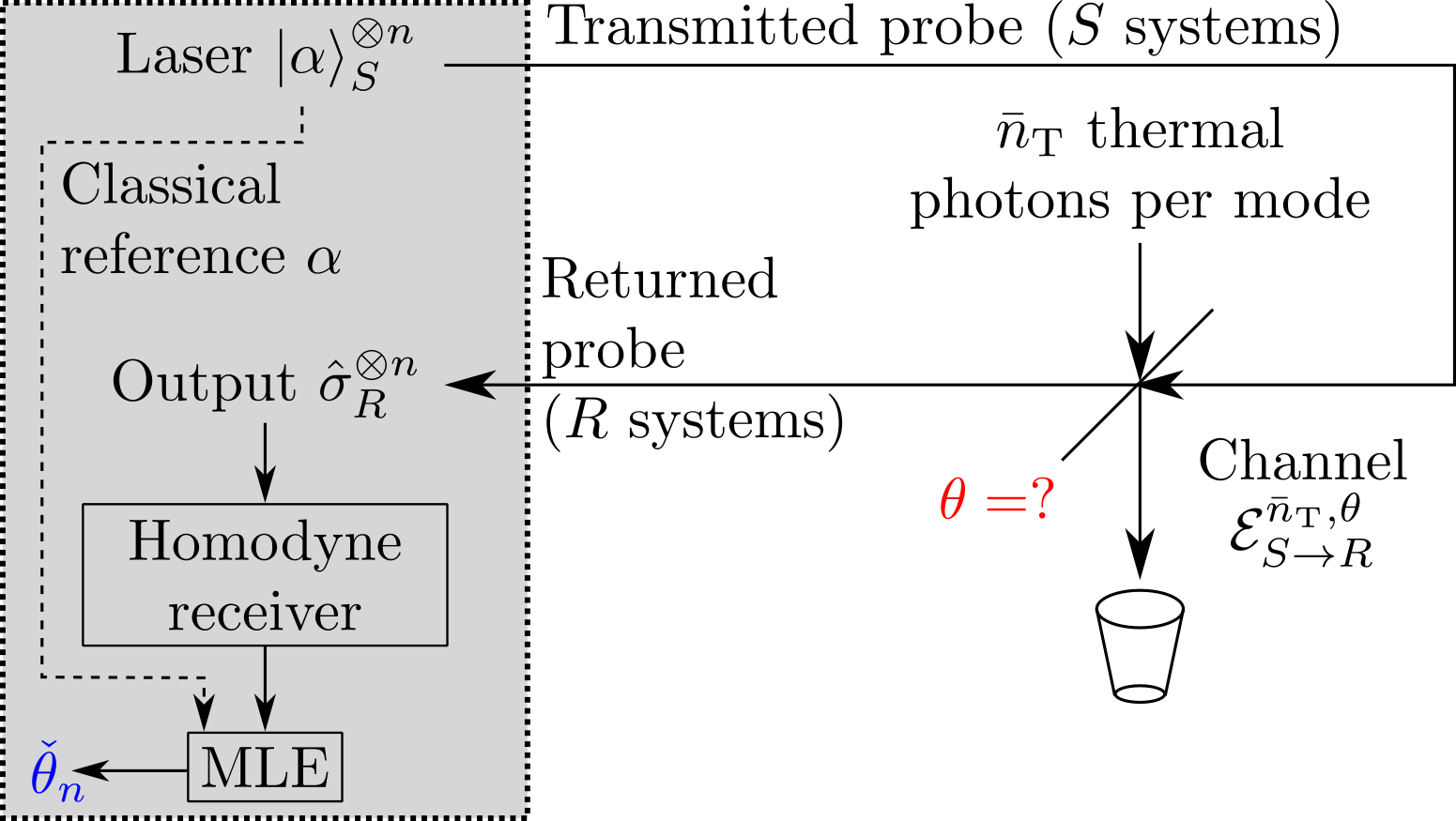}
	\caption{Sensing the unknown transmittance $\theta$ using $n$ coherent states $\ket{\alpha}_{S}$ and a homodyne receiver.  The output state $\hat{\sigma}_{R}$ is called a displaced thermal state. A Gaussian random variable describes the output of homodyne measurement \cite{weedbrook12gaussianQIrmp},\cite[Ch.~7.3.2]{guha04mastersthesis}.  A maximum likelihood estimator (MLE) that uses the homodyne receiver's output and the value of $\alpha$ as a classical reference achieves the quantum CRB for $\hat{\sigma}_{R}^{\otimes n}$ as $n\to\infty$, and is derived in Section \ref{sec:coherent_mle}.
 	 \label{fig:coherent_receiver}}
	\vspace{-0.1cm}
\end{figure}

Single-mode coherent state $\ket{\alpha}$ with complex-valued amplitude $\alpha\in\mathbb{C}$ is an eigenstate of the annihilation operator $\hat{a}$ and a quantum-mechanical description of laser light.
There are two types of coherent measurement: an optical homodyne receiver implements a measurement that uses the eigendecomposition of either $\hat{p}=\frac{\hat{a}-\hat{a}^\dagger}{\sqrt{2}}$ or $\hat{q}=\frac{\hat{a}+\hat{a}^\dagger}{\sqrt{2}}$ quadrature operators and yields a single Gaussian random variable, while a heterodyne receiver implements measurement that uses the eigendecompositions of both $\hat{p}$ and $\hat{q}$ and outputs a pair of independent Gaussian random variables.
A coherent transceiver employs both coherent states and a receiver, and is called ``classical'' in the literature.
The quantum FI associated with transmittance $\theta$ for a coherent state $\ket{\alpha}_{S}$ probe is \cite[Eq.~(23)]{nair20qi}:
\begin{align}
\label{eq:coh_qfi}\mathcal{J}_{\theta,\text{Coh}}\left(\bar{n}_{\rm S}\right)\equiv\mathcal{J}_{\theta}\left(\mathcal{E}^{\bar{n}_{\rm T},\theta}_{S\rightarrow R}\left[\ket{\alpha}\bra{\alpha}_{S}\right]\right)&=\frac{\bar{n}_{\rm S}}{\theta(2\bar{n}_{\rm B}+1)},
\end{align}
where the transmitted mean photon number $\bar{n}_{\rm S}\equiv|\alpha|^2$.
Use of a homodyne receiver as depicted in Fig.~\ref{fig:coherent_receiver} achieves the corresponding quantum CRB.

\subsection{TMSV and Optical Parametric Amplifier (OPA) Receiver}
\label{sec:OPA_receiver_description}
\begin{figure}[h]
\centering
	\includegraphics[width=0.8\columnwidth]{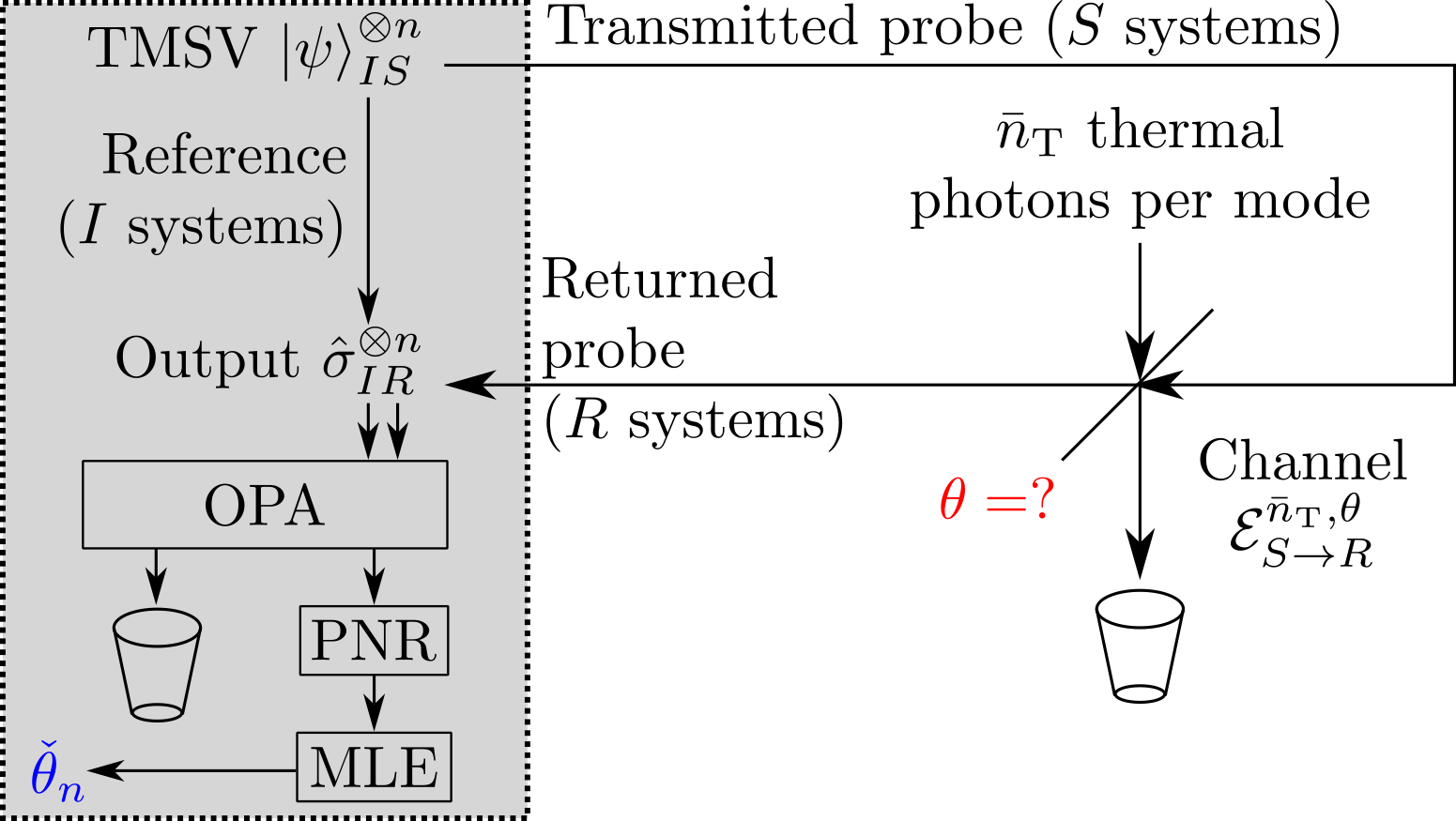}
	\caption{Sensing the unknown transmittance $\theta$ using $n$ two-mode squeezed vacuum (TMSV) states $\ket{\psi}_{IS}$ and an optical parametric amplifier (OPA) receiver.  Per \cite{guha09quantumilluminationOPA}, one of the outputs of the OPA is discarded, resulting in the other being in a thermal state.  A maximum likelihood estimator (MLE) that uses the output of the photon-number-resolving (PNR) measurement achieves the classical CRB asymptotically as $n\to\infty$, and is derived in Section \ref{sec:OPA_receiver}.
 	 \label{fig:OPA_receiver}}
	\vspace{-0.1cm}
\end{figure}

Let's use the TMSV probes as in Section \ref{sec:tmsv}, but apply an optical parametric amplifier (OPA) to the output state $\hat{\sigma}_{IR}(\theta)$ instead of a two-mode squeezer, and discard one of the OPA outputs.
The remaining output of the OPA is then in a zero-mean thermal state with average photon number \cite[Sec.~A]{guha09quantumilluminationOPA}:
\begin{align}
\nonumber \bar{n}_{\rm OPA}&=G\bar{n}_{\rm S}+(G-1)(\bar{n}_{\rm B}+1+\theta\bar{n}_{\rm S})\\
\label{eq:nbarOPA}&\phantom{=}+2\sqrt{G(G-1)\theta\bar{n}_{\rm S}(\bar{n}_{\rm S}+1)},
\end{align}
where the OPA gain $G>1$.
PNR measurement then outputs a random photon count $K$, distributed geometrically with mass function $q_K(k;\bar{n}_{\rm OPA})$ from \eqref{eq:tk}.
The OPA receiver was proposed for target detection with quantum illumination \cite{guha09quantumilluminationOPA, zhang15eesensing}, however, it can also be used to estimate transmittance $\theta$, as depicted in Fig.~\ref{fig:OPA_receiver}.
Classical FI associated with $\theta$ in $K$ is:
\begin{align}
\label{eq:OPA_cfi}\mathcal{J}_{\theta,\mathrm{OPA}}\left(\bar{n}_{\rm S}\right)&=\max_{G>1}\frac{\left((G-1)\bar{n}_{\rm S}+\sqrt{\frac{G(G-1)\bar{n}_{\rm S}(\bar{n}_{\rm S}+1)}{\theta}}\right)^2}{\bar{n}_{\rm OPA}(\bar{n}_{\rm OPA}+1)}.
\end{align}
For fixed $\bar{n}_{\rm S}>0$, the constrained maximization over gain $G$ is challenging analytically. 
Hence, we use numerical approaches to plot classical FI for an OPA receiver in Fig.~\ref{fig:qfi_plot}.
The dependence of optimal $G$ on $\theta$ can be addressed using the two-stage estimation approach described in Section \ref{sec:remarks_optimal_receiver}; we employ it in the Monte Carlo simulation described in Section \ref{sec:simulations}.
Furthermore, when $\bar{n}_{\rm S}$ is small, we have: 
\begin{align}
\label{eq:OPA_low_Ns}\left.\frac{\mathcal{J}_{\theta,\mathrm{OPA}}\left(\bar{n}_{\rm S}\right)}{\bar{n}_{\rm S}}\right|_{\bar{n}_{\rm S}=0}&=\max_{G>1}\frac{G}{\theta(\bar{n}_{\rm B}+1)(G(1+\bar{n}_{\rm B})-\bar{n}_{\rm B})}\\
&=\frac{1}{\theta(\bar{n}_{B}+1)},
\end{align}
where \eqref{eq:OPA_low_Ns} is maximized when $G\to1$.
This shows that the TMSV+OPA combination can perform close to the limit in \eqref{eq:ubQFI} in the low-transmitted-power low-transmittance regime. 

\subsection{Fock States}
\begin{figure}[h]
\centering
	\includegraphics[width=0.8\columnwidth]{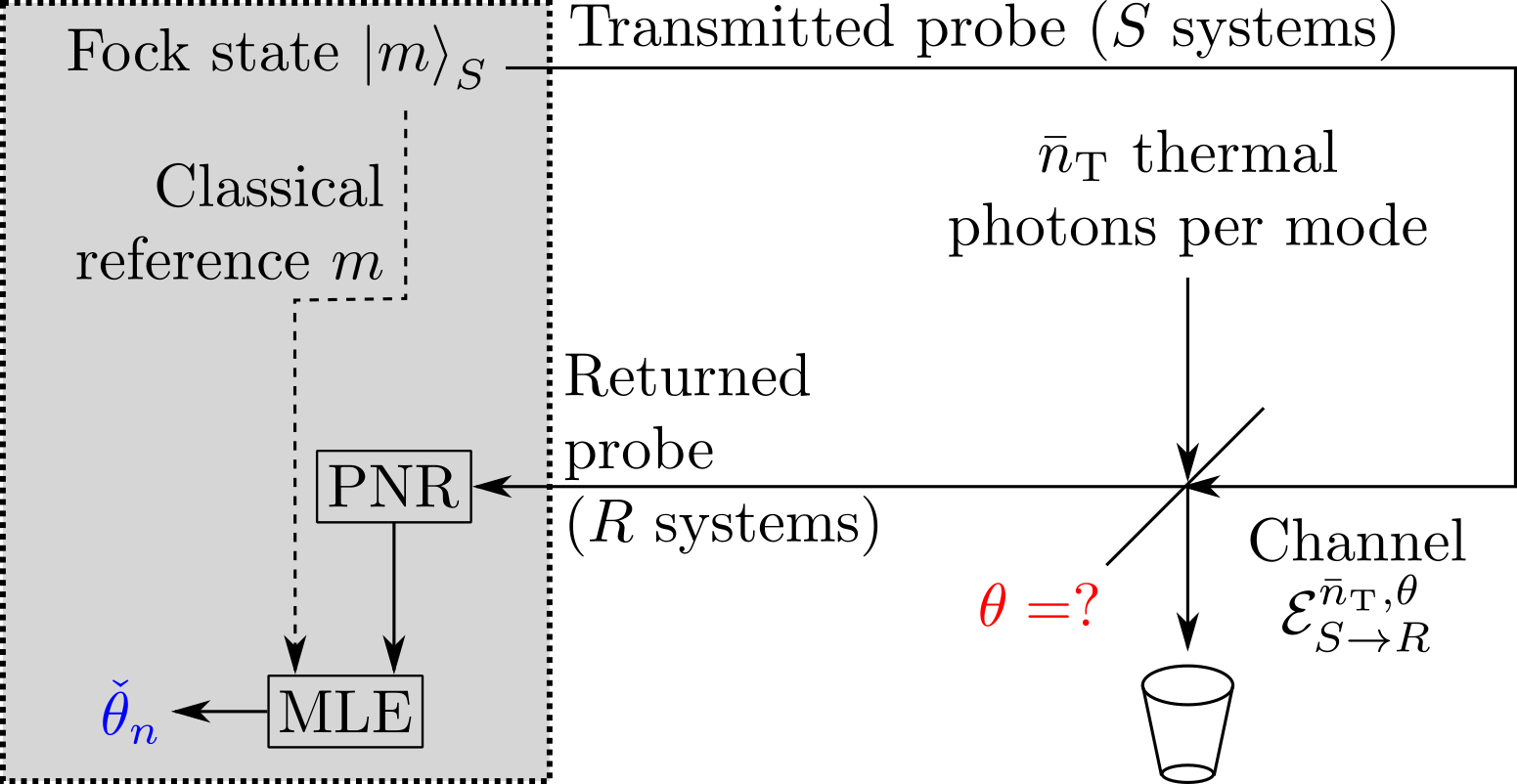}
	\caption{Sensing the unknown transmittance $\theta$ using $m$-photon Fock state $\ket{m}_{S}$.  Photon-number-resolving (PNR) measurement is optimal.  We can employ a maximum likelihood estimator (MLE) that uses the output of the PNR measurement and the value of $m$ as a classical reference. 
 	 \label{fig:Fock_receiver}}
	\vspace{-0.1cm}
\end{figure}

It is well-known that Fock state $\ket{m}$ is optimal for transmittance sensing in a vacuum ($\bar{n}_{\rm B}=0$), achieving quantum FI $\mathcal{J}_{\theta}\left(\mathcal{E}^{0,\theta}_{S\rightarrow R}\left[\ket{m}\bra{m}_{S}\right]\right)=\frac{m}{\theta(1-\theta)}$ \cite{adesso09optlossest}.
Here we show that this breaks down in thermal noise.
The output state $\hat{\nu}_R(\theta,m)=\mathcal{E}^{\bar{n}_{\rm T},\theta}_{S\rightarrow R}\left[\ket{m}\bra{m}_{S}\right]$ is diagonal in the Fock basis, making PNR measurement in the sensor as depicted in Fig.~\ref{fig:Fock_receiver} optimal.
The mass function for random variable $K$ describing the photon count is derived in \cite[Sec.~7.3.1]{guha04mastersthesis}.
Transformation of \cite[Eq.~(7.37)]{guha04mastersthesis} using \cite[9.131.1]{gr07tables} yields:
\begin{align}
\nonumber p_K(k;m)&=\binom{m+k}{k}\frac{(\bar{n}_{\rm B}+1-\theta)^{m-k}\theta^k}{(\bar{n}_{\rm B}+1)^{m+k+1}}\\
\label{eq:pnk_2F1trans}&\phantom{=}\times{}_2F_1\left[\begin{array}{c}-k,-k\\-(m+k)\end{array};z(\theta,\bar{n}_{\rm B})\right],
\end{align}
where $z(\theta,\bar{n}_{\rm B})=\frac{(\theta-\bar{n}_{\rm B})(\bar{n}_{\rm B}+1)}{\theta}$ and 
\begin{align}
\label{eq:2F1}{}_2F_1\left[\begin{array}{c}a,b\\ c\end{array};z\right]&=1+\frac{ab}{c}z+\frac{a(a+1)b(b+1)}{c(c+1)2!}z^2+\cdots
\end{align}
is the hypergeometric series.
The quantum FI associated with $\theta$ is:
\begin{align}
\nonumber\mathcal{J}_{\theta,\text{Fock}}(\ket{m})&\equiv\mathcal{J}_{\theta}\left(\hat{\nu}_R(\theta,m)\right)\\
\nonumber&=m\left(\frac{1-\theta}{(\bar{n}_{\rm B}+1-\theta)^2}+\frac{1}{\theta}\right)\\
\nonumber&\phantom{=}-\bar{n}_{\rm B}\left(\frac{1}{(\bar{n}_{\rm B}+1-\theta)^2}-\frac{1}{\theta^2}\right)\\
\nonumber&\phantom{=}-\frac{2\bar{n}_{\rm B}(\theta m+\bar{n}_{\rm B}+1)}{(\bar{n}_{\rm B}+1-\theta)\theta^2}\\
\nonumber&\phantom{=}-\frac{2\bar{n}_{\rm B}^2\left(2(\theta m+\bar{n}_{\rm B}+\theta m\bar{n}_{\rm B})+\bar{n}_{\rm B}^2+1\right)}{(\bar{n}_{\rm B}+1-\theta)^2\theta^2}\\
\label{eq:fock_qfi}&\phantom{=}-\frac{\bar{n}_{\rm B}^2m(m-1)}{(\bar{n}_{\rm B}+1-\theta)^2}+R_{m}(\theta,\bar{n}_{\rm B}),
\end{align}
where
\begin{align}
\nonumber\label{eq:R}R_{m}(\theta,\bar{n}_{\rm B})&=\sum_{k=0}^\infty \binom{m+k}{k}\frac{k^4\bar{n}_{\rm B}^2(\bar{n}_{\rm B}+1-\theta)^{m-k}\theta^{k-4}}{(m+k)^2(\bar{n}_{\rm B}+1)^{m+k-1}}\\
&\phantom{=\sum_{k=0}^\infty}\times\frac{{}_2F_1\left[\begin{array}{c}-(k-1),-(k-1)\\-(m+k-1)\end{array};z(\theta,\bar{n}_{\rm B})\right]^2}{{}_2F_1\left[\begin{array}{c}-k,-k\\-(m+k)\end{array};z(\theta,\bar{n}_{\rm B})\right]}.
\end{align}
The ratio of the hypergeometric series in \eqref{eq:R} makes both the analysis and numerical evaluation of the quantum FI $\mathcal{J}_{\theta,\text{Fock}}(\ket{m})$ for large $m$ challenging.
Furthermore, on-demand generation of Fock states $\ket{m}$ for arbitrary $m$  presents technical challenges that appear insurmountable in the near term.
However, such sources exist for single photons \cite{eisaman11singlphotons,sinha19singlephotonsources,meyer-scott20singlephotonsources}.
Setting $m=1$ in \eqref{eq:pnk_2F1trans} yields 
\begin{align}
\label{eq:p1k}p_K(k;1)&=\left(\bar{n}_{\rm B}(\bar{n}_{\rm B}+1)+\theta(k-\bar{n}_{\rm B})\right)\bar{n}_{\rm B}^{k-1}(\bar{n}_{\rm B}+1)^{-(k+2)}.
\end{align}
The corresponding quantum FI is
\begin{align}
\label{eq:singlephoton_qfi}\mathcal{J}_{\theta,\text{Fock}}\left(\ket{1}\right)&=\sum_{k=0}^\infty \frac{(k-\bar{n}_{\rm B})^2\bar{n}_{\rm B}^{k-1}(\bar{n}_{\rm B}+1)^{-(k+2)}}{\left(\bar{n}_{\rm B}(\bar{n}_{\rm B}+1)+\theta(k-\bar{n}_{\rm B})\right)}.
\end{align}
We note that, for $\bar{n}_{\rm B}>0$, \eqref{eq:singlephoton_qfi} is well approximated by the first three terms of the sum.
The contribution from terms corresponding to $k>2$ decreases rapidly due to the exponential decay of $\bar{n}_{\rm B}^{k-1}(\bar{n}_{\rm B}+1)^{-(k+2)}$ with increasing $k$.

\subsection{TMSV and Heralded PNR Measurement}
\label{sec:heralded_receiver}
\begin{figure}[h]
\centering
	\includegraphics[width=0.8\columnwidth]{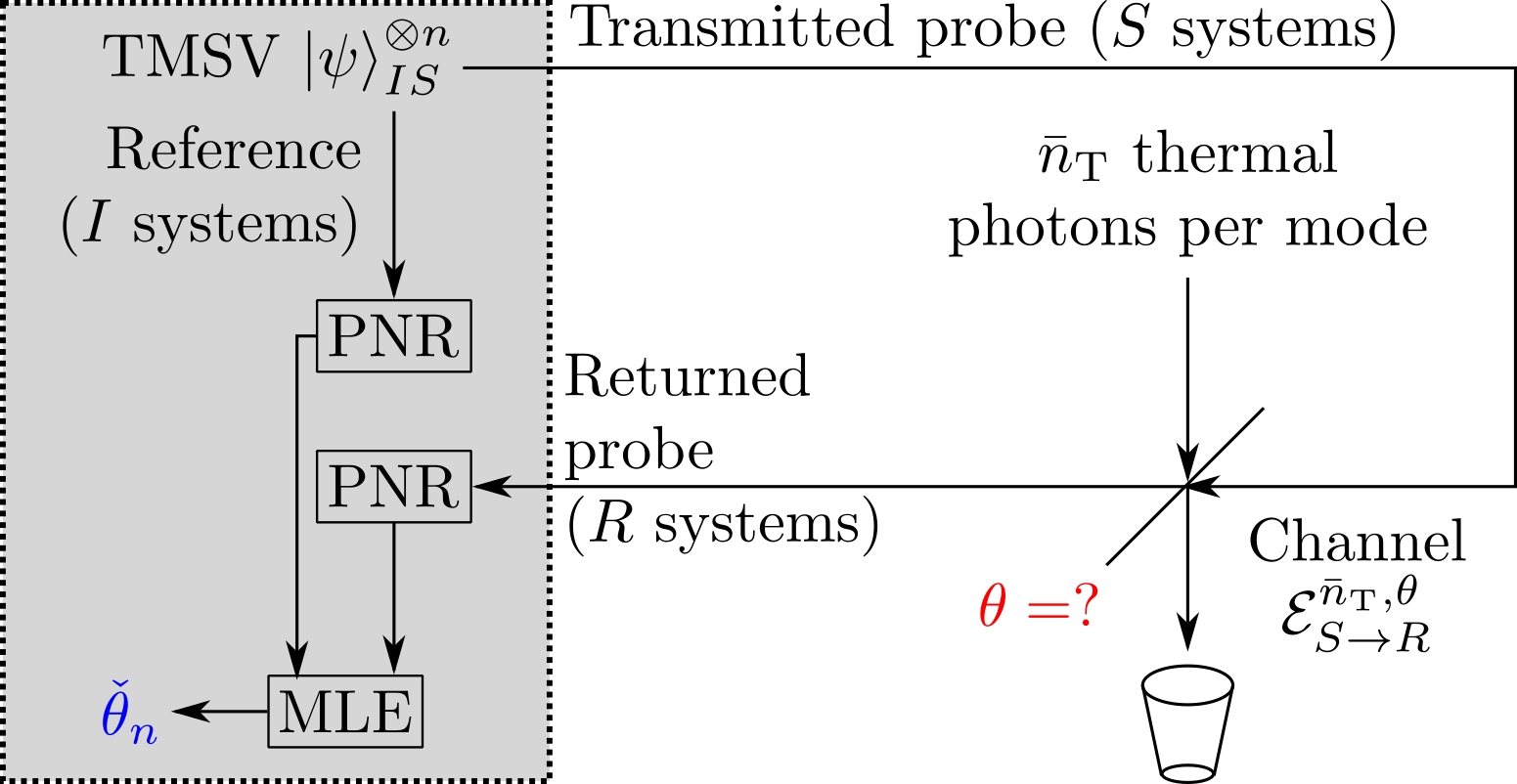}
	\caption{Sensing the unknown transmittance $\theta$ using $n$ two-mode squeezed vacuum (TMSV) states $\ket{\psi}_{IS}$.  The  photon-number-resolving (PNR) measurement of the idler mode heralds the photon number of the transmitted Fock state probe.  This is used by a maximum likelihood estimator (MLE) along with the output of the PNR measurement of the returned probe to achieve the classical CRB asymptotically as $n\to\infty$.
 	 \label{fig:heralded_receiver}}
	\vspace{-0.1cm}
\end{figure}

TMSV states and PNR measurement can be used for probabilistic generation of Fock states.
Detection of $m$ photons on the idler mode heralds $\ket{m}_S$ state on the signal mode.
Thus, we consider transmittance sensing with TMSV using two PNR measurements: one for the idler mode, and the other for the returned signal port.
This is depicted in Fig.~\ref{fig:heralded_receiver}.
The classical FI of this system is 
\begin{align}
\label{eq:her_cfi}\mathcal{J}_{\theta,\mathrm{Her}}\left(\bar{n}_{\rm S}\right)&=E_{M}\left[\mathcal{J}_{\theta,\text{Fock}}(\ket{M})\right],
\end{align}
where the output photon number $M$ mass function is in \eqref{eq:tk}.

\subsection{Comparison}

\begin{figure*}
\centering
\subfloat[$\bar{n}_{\rm S}=0.001$]{\includegraphics[width=0.47\linewidth]{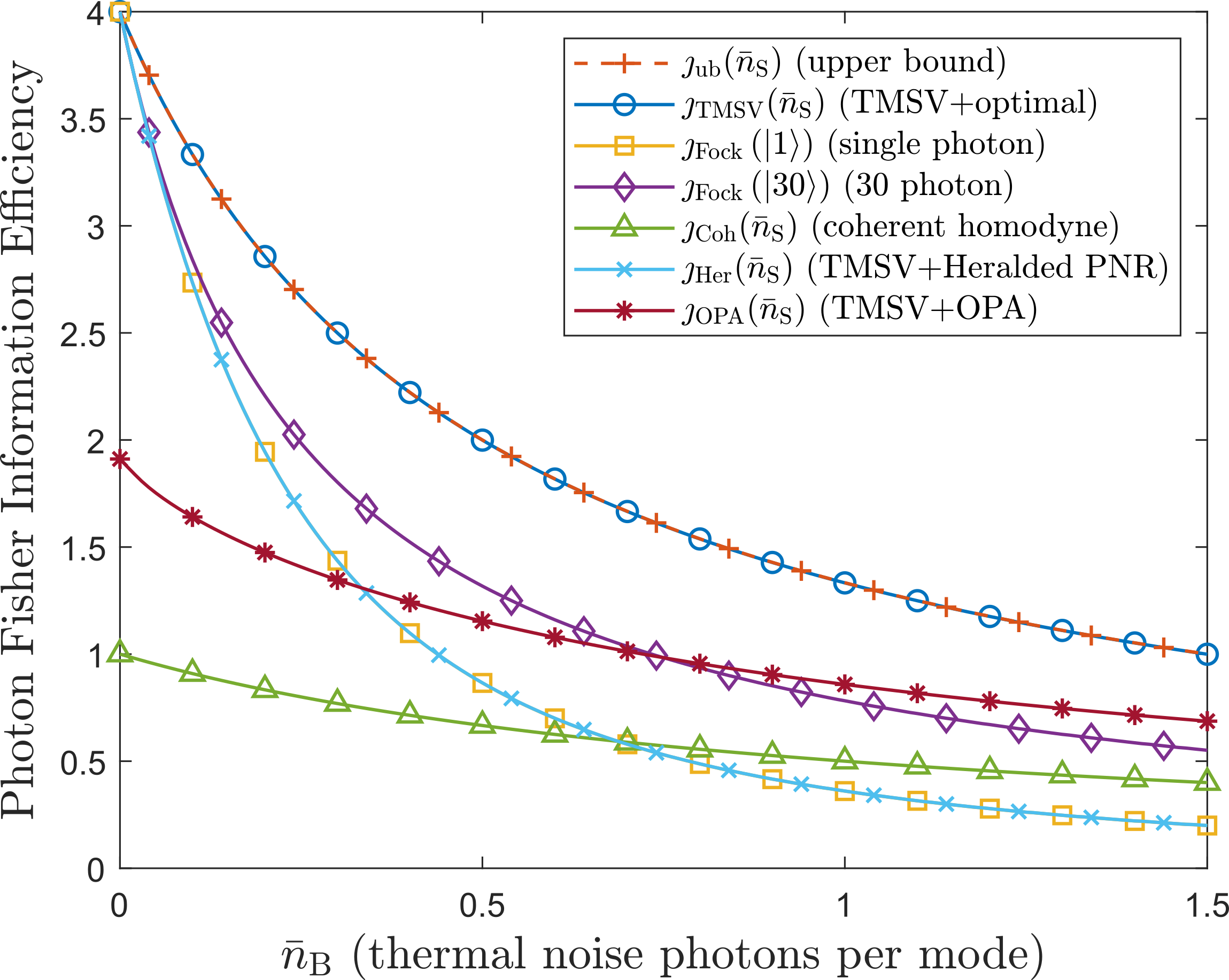}}
\hfil
\subfloat[$\bar{n}_{\rm S}=0.01$]{\includegraphics[width=0.47\linewidth]{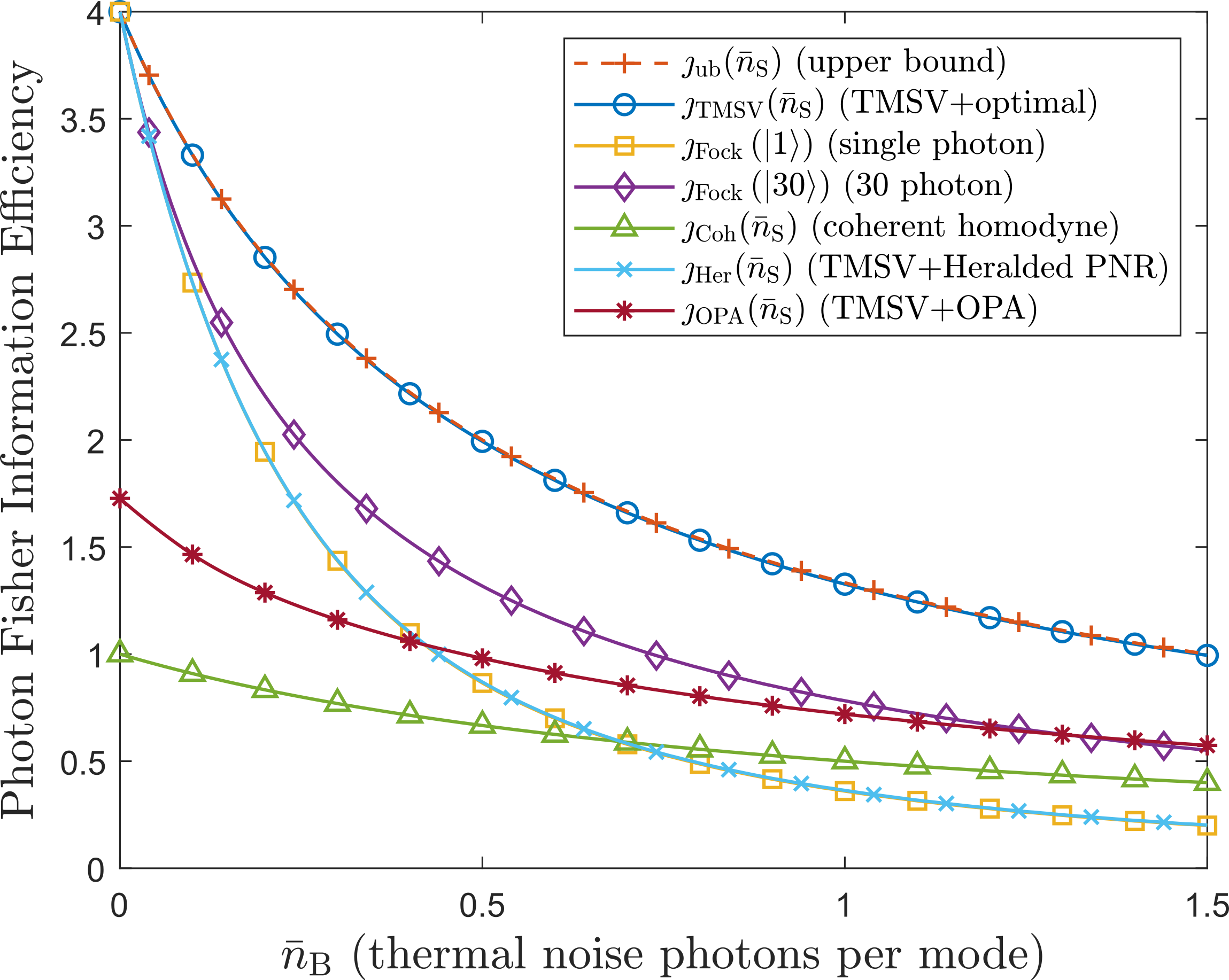}}\\
\subfloat[$\bar{n}_{\rm S}=0.1$]{\includegraphics[width=0.47\linewidth]{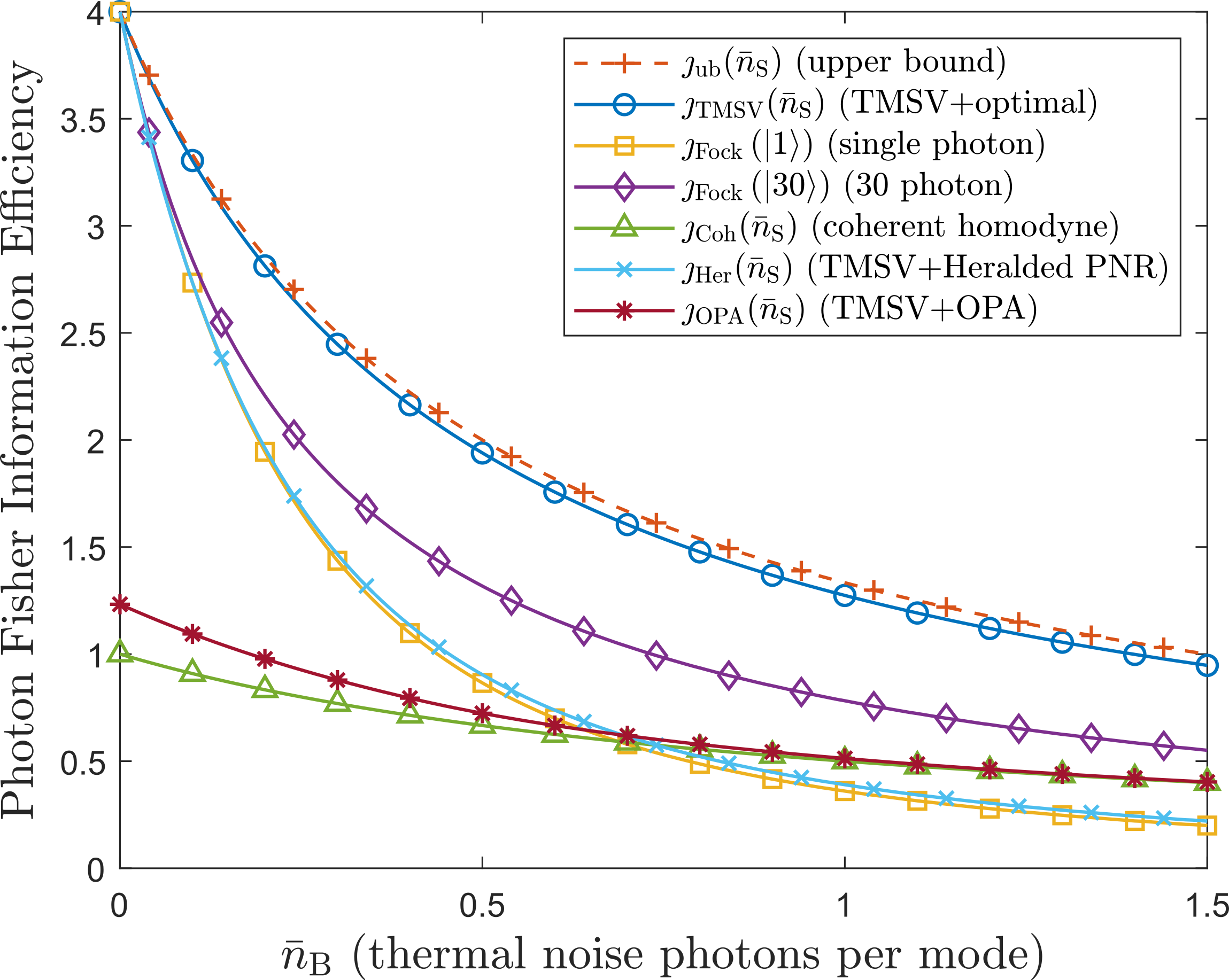}}
\hfil
\subfloat[$\bar{n}_{\rm S}=1$]{\includegraphics[width=0.47\linewidth]{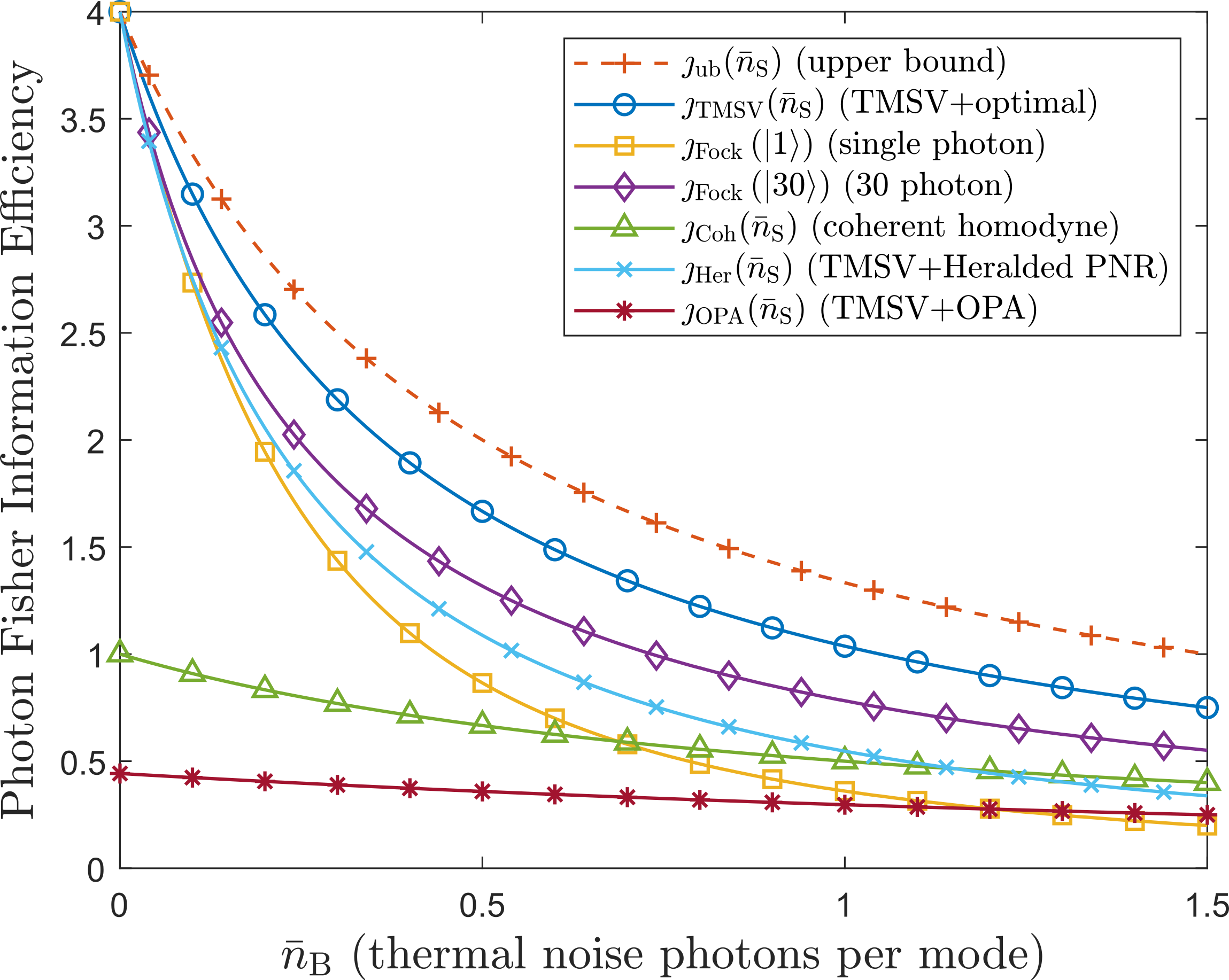}}\\
\subfloat[$\bar{n}_{\rm S}=10$]{\includegraphics[width=0.47\linewidth]{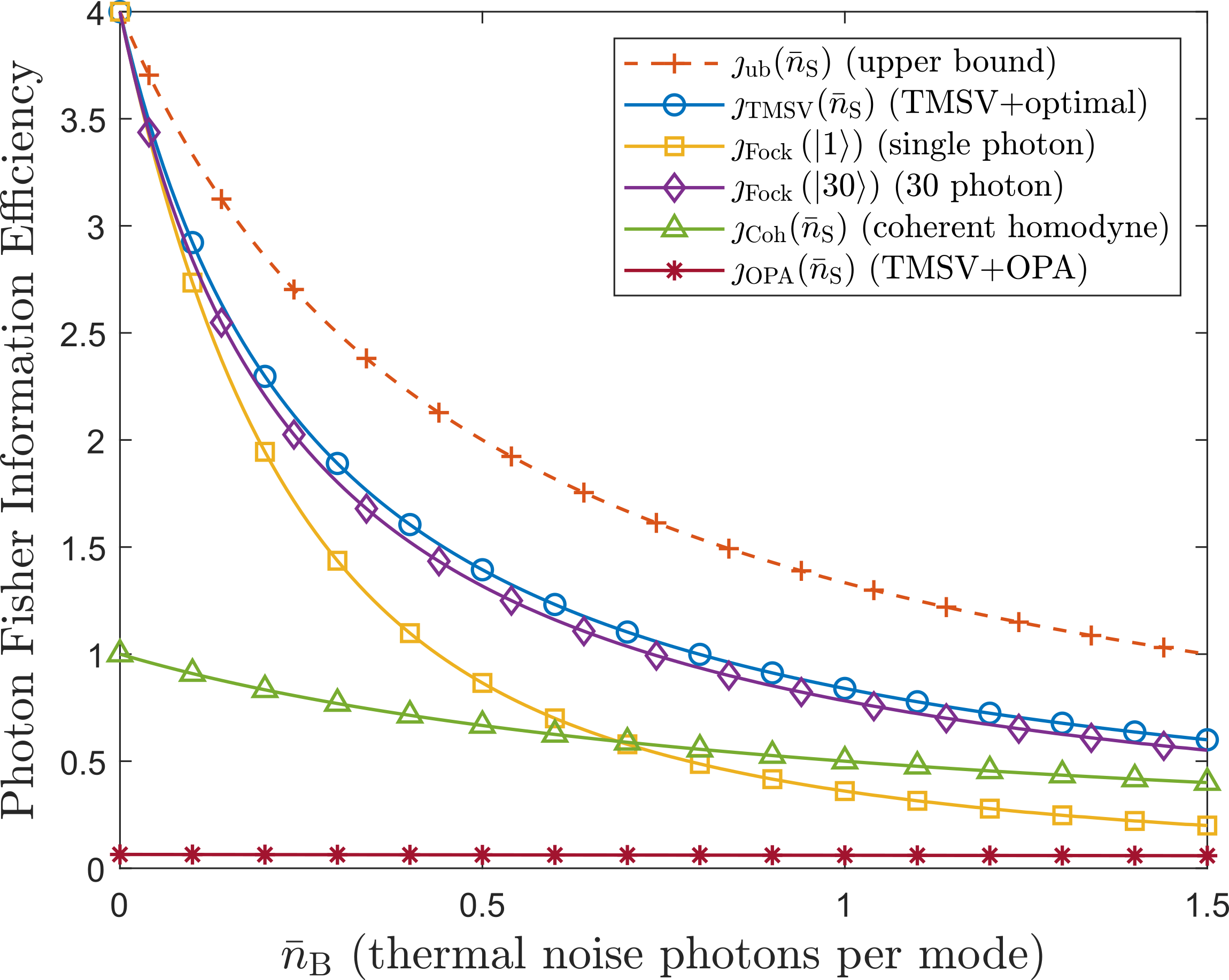}}
\hfil
\subfloat[$\bar{n}_{\rm S}=100$]{\includegraphics[width=0.47\linewidth]{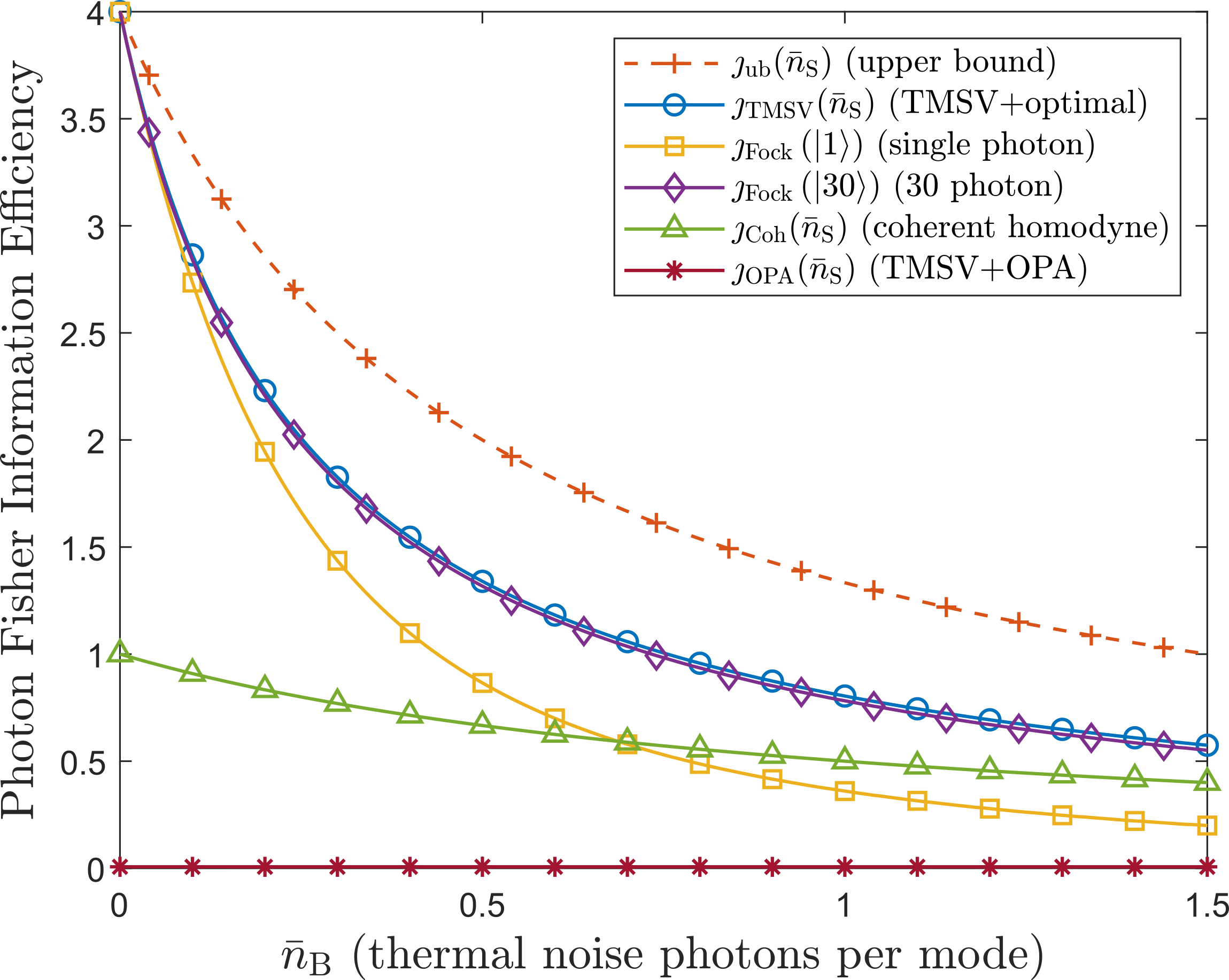}}
\caption{Photon Fisher information efficiency vs.~thermal-noise mean number of photons per mode $\bar{n}_{\rm B}$ for various transmittance estimation methods and values of transmitted mean  number of photons per mode $\bar{n}_{\rm S}$.  We set $\theta=0.5$, results for other values of $\theta$ are qualitatively similar.}
\label{fig:qfi_plot}
\end{figure*}

We ensure a fair comparison of the fundamental limits for various transmittance-sensing methods by analyzing their Fisher information attained per photon per (transmitted) mode. That is, \emph{photon Fisher information efficiency} (PFIE) $\jmath\left(\bar{n}_{\rm S}\right)\equiv\frac{\mathcal{J}_\theta\left(\bar{n}_{\rm S}\right)}{\bar{n}_{\rm S}}$ is our figure of merit, where $\mathcal{J}_\theta\left(\bar{n}_{\rm S}\right)$ is classical or quantum FI attainable with $\bar{n}_{\rm S}$ input mean signal photons per mode.
The ultimate upper bound for PFIE is derived from \eqref{eq:ubQFI}: 
\begin{align}
\label{eq:ubPFIE}\jmath_{\rm ub}\left(\bar{n}_{\rm S}\right)&\equiv\frac{\mathcal{J}_{\theta,\rm{ub}}\left(\bar{n}_{\rm S}\right)}{\bar{n}_{\rm S}}=\frac{1}{\theta(\bar{n}_{\rm B}+1-\theta)}.
\end{align}
The PFIE for TMSV source uses the quantum FI in \eqref{eq:tmsv_qfi}:
\begin{align}
\nonumber\jmath_{\rm TMSV}\left(\bar{n}_{\rm S}\right)&\equiv\frac{\mathcal{J}_{\theta,\text{TMSV}}\left(\bar{n}_{\rm S}\right)}{\bar{n}_{\rm S}}\\
\label{eq:tmsv_pfie}&=\frac{\bar{n}_{\rm B}+1+(1-\theta)\bar{n}_{\rm S}+\bar{n}_{\rm B}\bar{n}_{\rm S}}{\theta\left(\bar{n}_{\rm B}+1-\theta\right)\left(\bar{n}_{\rm B}+1+\bar{n}_{\rm S}(2\bar{n}_{\rm B}+1-\theta)\right)}.
\end{align}
The PFIE for coherent and TMSV+OPA transceivers use classical FI expressions \eqref{eq:coh_qfi} and \eqref{eq:OPA_cfi}, respectively:
\begin{align}
\label{eq:coh_pfie}\jmath_{\rm Coh}\left(\bar{n}_{\rm S}\right)&\equiv\frac{\mathcal{J}_{\theta,\text{Coh}}\left(\bar{n}_{\rm S}\right)}{\bar{n}_{\rm S}}=\frac{1}{2\theta(\bar{n}_{\rm B}+1)}\\
\nonumber\jmath_{\rm OPA}\left(\bar{n}_{\rm S}\right)&\equiv\frac{\mathcal{J}_{\theta,\mathrm{OPA}}\left(\bar{n}_{\rm S}\right)}{\bar{n}_{\rm S}}\\
\label{eq:pfie_opa}&=\max_{G>1}\frac{\left((G-1)\bar{n}_{\rm S}+\sqrt{\frac{G(G-1)\bar{n}_{\rm S}(\bar{n}_{\rm S}+1)}{\theta}}\right)^2}{\bar{n}_{\rm S}\bar{n}_{\rm OPA}(\bar{n}_{\rm OPA}+1)}.
\end{align}
The PFIE for Fock state and TMSV+Heralded PNR measurement use quantum and classical FI expressions in \eqref{eq:fock_qfi} and \eqref{eq:her_cfi}, respectively:
\begin{align}
\label{eq:fock_pfie}\jmath_{\rm Fock}\left(\ket{m}\right)&\equiv \frac{\mathcal{J}_{\theta,\text{Fock}}\left(\ket{m}\right)}{m}\\
\label{eq:her_pfie}\jmath_{\rm Her}\left(\bar{n}_{\rm S}\right)&\equiv \frac{\mathcal{J}_{\theta,\mathrm{Her}}\left(\bar{n}_{\rm S}\right)}{\bar{n}_{\rm S}}.
\end{align} 
We evaluate \eqref{eq:ubPFIE}-\eqref{eq:her_pfie} and plot the results versus thermal-noise mean photon number $\bar{n}_{\rm B}$ in Fig.~\ref{fig:qfi_plot}.
While $\jmath_{\rm ub}\left(\bar{n}_{\rm S}\right)$ and $\jmath_{\rm Coh}\left(\bar{n}_{\rm S}\right)$ are constant relative to transmitted mean photon number $\bar{n}_{\rm S}$, $\jmath_{\rm TMSV}\left(\bar{n}_{\rm S}\right)$, $\jmath_{\rm OPA}\left(\bar{n}_{\rm S}\right)$, and $\jmath_{\rm Her}\left(\bar{n}_{\rm S}\right)$ are not.
Thus, we include plots for various values of $\bar{n}_{\rm S}$.
We set $\theta=0.5$ and note that, for other values $\theta$, the plots are qualitatively similar.
In plotting $\jmath_{\rm OPA}\left(\bar{n}_{\rm S}\right)$ we maximize \eqref{eq:pfie_opa} over $G>1$ numerically.
We include the plot of $\jmath_{\rm Fock}\left(\ket{m}\right)$ for single-photon Fock state $\ket{1}$ and a thirty-photon Fock state $\ket{30}$ on all plots.
The challenges associated with computing the hypergeometric series ${}_2F_1$ \cite{pearson17hypergeometric} precluded evaluating $\jmath_{\rm Fock}\left(\ket{m}\right)$ for $m>30$.
When evaluating $\jmath_{\rm Her}\left(\bar{n}_{\rm S}\right)$, we had to truncate the sum in the expectation in \eqref{eq:her_cfi} at $m=30$.
This accurately approximates $\jmath_{\rm Her}\left(\bar{n}_{\rm S}\right)$ only up to $\bar{n}_{\rm S}\approx 4$.
Thus, we did not evaluate $\left.\jmath_{\rm Her}\left(\bar{n}_{\rm S}\right)\right|_{\bar{n}_{\rm S}=10}$ and $\left.\jmath_{\rm Her}\left(\bar{n}_{\rm S}\right)\right|_{\bar{n}_{\rm S}=100}$.

It is evident from Fig.~\ref{fig:qfi_plot} that PFIE of the TMSV input combined with optimal measurement exceeds that of other receivers we consider. 
While Fock state transmitters' performance rapidly decays with noise, they outperform the TMSV+OPA transceiver when signal-to-noise ratio (SNR) is high.
Indeed, the TMSV+OPA transceiver performs very poorly when the transmitted mean photon number is high.

On the other hand, the high PFIE of single-photon Fock state $\ket{1}$ in low noise shows the promise of using the on-demand single-photon sources and PNR measurement for transmittance sensing.
Furthermore, since the first three terms of the summation in \eqref{eq:singlephoton_qfi} approximate $\mathcal{J}_{\theta}\left(\hat{\nu}_R(\theta,1)\right)$ well for $\bar{n}_{\rm B}<1$, a measurement that distinguishes zero, one, or more photons suffices for accurate estimation of $\theta$ for these systems.
Such measurement is less complex than the full PNR one.

We also note that TMSV+heralded PNR performs as well as the single-photon Fock state source in low-noise setting.
In fact, our calculations suggest that, for $\bar{n}_{\rm S}\ll 1$, TMSV+heralded PNR measurement matches PFIE of a single-photon Fock state source while using a less-complex single-photon detector (SPD) that distinguishes zero or more than zero photons instead of a PNR measurement.

Nevertheless, many practical scenarios demand low-transmitted-power operation.  
Even for moderate noise power, this results in low SNR.
Transmittance sensors that employ TMSV+receiver derived in Section \ref{sec:optimal_receiver} and TMSV+OPA receiver behave well in this setting.
Thus, next we use Monte Carlo simulation to evaluate these sensors and to compare their performance to a simpler one based of a coherent transceiver.

\section{Comparison with Alternative Transmittance-Sensing Methods: Simulations}
\label{sec:simulations}

Maximum likelihood estimators (MLEs) have a number of desirable properties, the first and foremost being the availability of ``turn-the-crank'' implementation in most practical settings.  Furthermore, MLEs are usually asymptotically consistent and efficient, as the number of observations $n\to\infty$ \cite{vantrees01part1,kay93statSP1}.  Here we employ MLEs to estimate transmittance $\theta$ from the outputs of coherent homodyne transceiver, TMSV+OPA receiver, and TMSV+receiver derived in Section \ref{sec:optimal_receiver}, analyzing their convergence to CRB for increasing $n$.

\subsection{Construction of MLEs}
\label{sec:mle}

\subsubsection{Coherent Homodyne Transceiver}
\label{sec:coherent_mle}

Consider a transmittance sensing scheme described in Section \ref{sec:coherent_transceiver} that uses a tensor product $\ket{\alpha}^{\otimes n}_{S}$ of $n$ coherent states $\ket{\alpha}_S$ with $\alpha=\sqrt{\bar{n}_{\rm S}}$ as probes and a homodyne receiver.
The corresponding output is a sequence of $n$ independent and identically distributed (i.i.d.) Gaussian random variables $\left\{X_k\right\}_{k=1}^n$, each with mean $\sqrt{\theta\bar{n}_{\rm S}}$ and variance $\bar{n}_{\rm B}+1$.  The MLE for $\theta$ is:
\begin{align}
\check{\theta}_{\text{Coh}}(n)&=\frac{1}{\bar{n}_{\rm S}}\left(\frac{1}{n}\sum_{k=1}^n x_k\right)^2,
\end{align}
where $x_k$ is an observed instance of $X_k$.

\subsubsection{TMSV Input and OPA Receiver}
\label{sec:OPA_receiver}
Now consider a scheme from Section \ref{sec:OPA_receiver_description} that uses a tensor product $\hat{\rho}_{I^nS^n}=\ket{\psi}\bra{\psi}_{IS}^{\otimes n}$ of $n$ TMSV states defined in \eqref{eq:tmsv} as probes and an OPA receiver.
The corresponding output is a sequence of $n$ i.i.d.~geometric random variables $\left\{Y_k\right\}_{k=1}^n$, each with mass function $q_Y(y;\bar{n}_{\rm OPA})$ defined in \eqref{eq:tk}, where the $\bar{n}_{\rm OPA}$ is in \eqref{eq:nbarOPA}.  The MLE for $\theta$ is:
\begin{align}
\nonumber\check{\theta}_{\rm OPA}(n)&=\frac{\bar{y}(G^{\ast}-1)-1-\bar{n}_{\rm B}}{(G^{\ast}-1)^2\bar{n}_{\rm S}}\\
\nonumber&\phantom{=}-\frac{2\sqrt{G^{\ast}(1+\bar{y}+\bar{n}_{\rm B}-G^{\ast}\bar{n}_{\rm S})(1+\bar{n}_{\rm S})}}{(G^{\ast}-1)\bar{n}_{\rm S}}\\
\label{eq:OPA_mle}&\phantom{=}+\frac{G^{\ast}(G^{\ast}+2\bar{n}_{\rm B}-G^{\ast}\bar{n}_{\rm B}+(G^{\ast}-1)\bar{n}_{\rm S})}{(G^{\ast}-1)^2\bar{n}_{\rm S}},
\end{align}
where $\bar{y}=\frac{1}{n}\sum_{k=1}^n y_k$ is the mean of the observed instances of $Y_k$, and $G^{\ast}$ is the OPA gain that maximizes classical FI in \eqref{eq:OPA_cfi}.
Thus, $G^{\ast}$ minimizes the CRB for the receiver, and hence, the asymptotic MSE of $\check{\theta}_{\rm OPA}(n)$.
However, $G^{\ast}$ depends on the parameter of interest, $\theta$.
Thus, we follow a two-stage approach \cite[Ch.~6.4]{hayashi17qit},  \cite{gill00twostagemeasurement,hayashi05twostagemeasurement}.  First $\sqrt{n}$ probes are coherent states from which we obtain a preliminary estimate $\check{\theta}_0=\check{\theta}_{\rm Coh}\left(\sqrt{n}\right)$, using the homodyne receiver described above. 
We compute $G^{\ast}_0$ by maximizing \eqref{eq:OPA_cfi} with $\check{\theta}_0$ substituted for $\theta$.
The remaining $n-\sqrt{n}$ probes are TMSV states, with output processed by the OPA receiver, with gain set to $G^{\ast}_0$, and the corresponding MLE $\check{\theta}_{\rm OPA}\left(n-\sqrt{n}\right)$ in \eqref{eq:OPA_mle}, with $G^{\ast}$ set to $G^{\ast}_0$.

\subsubsection{TMSV Input and Receiver Derived in Section \ref{sec:optimal_receiver}}
Finally, we use a tensor product $\hat{\rho}_{I^nS^n}=\ket{\psi}\bra{\psi}_{IS}^{\otimes n}$ of $n$ TMSV states defined in \eqref{eq:tmsv} as probes, but we employ the receiver derived in Section \ref{sec:optimal_receiver}, that uses the two-mode squeezer $\hat{S}(\omega)$, followed by the PNR measurement of each mode.
As noted in Section \ref{sec:remarks_optimal_receiver}, the squeezing parameter $\omega$ depends on the parameter of interest $\theta$.
We follow a two-stage approach as in the OPA-based scheme, with an identical first stage that uses a coherent homodyne transceiver.
We calculate the squeezing parameter $\omega_0$ using the preliminary estimate $\check{\theta}_0$ that employs the first $\sqrt{n}$ probes.
The remaining $n-\sqrt{n}$ probes are TMSV states, with each output state $\hat{\sigma}_{IR}(\theta)$.
These are processed by $\hat{S}\left(\omega_0\right)$ followed by a two-mode PNR measurement.
Thus, the corresponding output is a sequence of $n$ i.i.d.~pairs of random variables $\left\{\left(Z_{0,k},Z_{1,k}\right)\right\}_{k=1}^n$, each with mass function:
\begin{align}
\label{eq:tmsv_opt_p}p_{Z_0,Z_1}(z_0,z_1)&=\bra{z_0z_1}\hat{S}^{\dagger}\left(\omega_0\right) \hat{\sigma}_{IR}(\theta) \hat{S}\left(\omega_0\right) \ket{z_0z_1}\\
\label{eq:tmsv_opt_p1}&=\sum_{s=0}^\infty\sum_{t=0}^\infty r_{st} \left| \bra{st} \hat{S}\left(\omega_0+\zeta\right) \ket{z_0z_1} \right|^2,
\end{align}
where, in Appendix \ref{app:focksqueezing}, we derive
\begin{multline}
   \bra{st} S\left(\omega_0+\zeta\right)\ket{km}\\
   =(-\tau_0)^{s-k}\sqrt{s!t!k!m!}\delta(s-k,t-m)\qquad\\
   \phantom{=}\times\sum_{u=u_{\rm l}}^{u_{\rm m}}\frac{(-|\tau_0|^2)^u \nu_0^{-k-m+2u - 1}}{(s-k+u)!u!(k-u)!(m-u)!}\label{eq:focksqueezing}
\end{multline}
with $u_{\rm l}=\max(k-s,0)$, $u_{\rm m}=\min(k,m)$, $\tau_0=\frac{\omega_0+\zeta}{|\omega_0+\zeta|}\tanh|\omega_0+\zeta|$, $\nu_0=\cosh|\omega_0+\zeta|$, the Kronecker delta function $\delta(x,y)=\{1\text{~if~}x=y; 0\text{~if~}x\neq y\}$, and $\zeta$ and $r_{st}$ defined in \eqref{eq:zeta} and \eqref{eq:rkm}, respectively.
Another form of \eqref{eq:focksqueezing} is in \cite[Eq.~(22)]{hong-yi96twomodesqueezer}.
Since we multiply by $\delta(s-k,t-m)$, \eqref{eq:focksqueezing} is not zero only if $s-k=t-m$.
Thus, \eqref{eq:tmsv_opt_p1} simplifies to:
\begin{align}
\label{eq:tmsv_opt_p2}p_{Z_0,Z_1}(z_0,z_1)&=\sum_{s=0}^\infty r_{st'} \left| \bra{st'} \hat{S}\left(\omega_0+\zeta\right) \ket{z_0z_1} \right|^2,
\end{align}
where $t'=s-k+m$.
As there is no known closed-form solution for MLE, we construct it using numerical optimization of \eqref{eq:tmsv_opt_p2} recast as a likelihood function for $\theta$.

\subsection{Results} 
\label{sec:results}
While it is well-known that MLE is asymptotically consistent and efficient, as number of observations $n\to\infty$ \cite{vantrees01part1,kay93statSP1}, numerical approaches are needed for its \emph{finite-sample} performance analysis at fixed $n<\infty$.
Furthermore, although the two-stage approach \cite[Ch.~6.4]{hayashi17qit},  \cite{gill00twostagemeasurement,hayashi05twostagemeasurement} is consistent and quantum efficient as $n\to\infty$, the convergence conditions in \cite[Ch.~6.4]{hayashi17qit},  \cite{hayashi05twostagemeasurement} are onerous to prove mathematically for many estimators (including ours) and, to our knowledge, its finite sample analysis is missing from the literature.

Thus, we compare the simulated MSEs for the MLEs developed in Section \ref{sec:mle} to their corresponding CRBs.
Our settings for $\bar{n}_{\rm S}=0.01$, $\bar{n}_{\rm B}=1$, and $\theta=0.5$ are chosen because 1) they allow the simulations to complete in reasonable time while ensuring the existence of the receiver derived in Section \ref{sec:optimal_receiver}; and, 2) $\bar{n}_{\rm S}\ll\bar{n}_{\rm B}$ models transmittance sensing in a low-SNR environment.
The MSE is inversely proportional to the number of probes $n$: $\text{MSE}_{\theta}=\frac{c_{\theta}}{n}$.
For $\bar{n}_{\rm S}$, $\bar{n}_{\rm B}$, and $\theta$ fixed, the scaling factor $c_{\theta}\equiv n\times\text{MSE}_{\theta}$ is a function of $n$, however, the asymptotic efficiency of MLE suggests that, for estimators described in Section \ref{sec:mle}, $\lim_{n\to\infty}c_{\theta}=\frac{1}{\mathcal{J}_{\theta}}$, where $\frac{1}{\mathcal{J}_{\theta}}$ is a single-observation CRB.\footnote{As $n\to\infty$, MLE $\check{\theta}_{\rm MLE}(n)$ for $\theta$ is usually asymptotically efficient: $\check{\theta}_{\rm MLE}(n)$ converges in probability to the true value $\theta_0$. It is also typically asymptotically efficient as follows: $\sqrt{n}\left(\check{\theta}_{\rm MLE}(n)-\theta_0\right)$ converges in law to a zero-mean Gaussian random variable with variance equal to CRB for a single observation \cite{vantrees01part1,kay93statSP1}.  However, proving $\lim_{n\to\infty}c_{\theta}=\frac{1}{\mathcal{J}_{\theta}}$ mathematically requires showing uniform integrability, which is challenging for MLEs (see, e.g., remarks following Proposition IV.D.2 in \cite[Sec.~IV.D]{poor94intro-det-est}).}
We are interested in the speed of $c_{\theta}$'s convergence to $\frac{1}{\mathcal{J}_{\theta}}$ as $n$ increases, and the penalty (if any) of the two-stage approach.

\begin{figure}[h]
\centering
	\includegraphics[width=1\columnwidth]{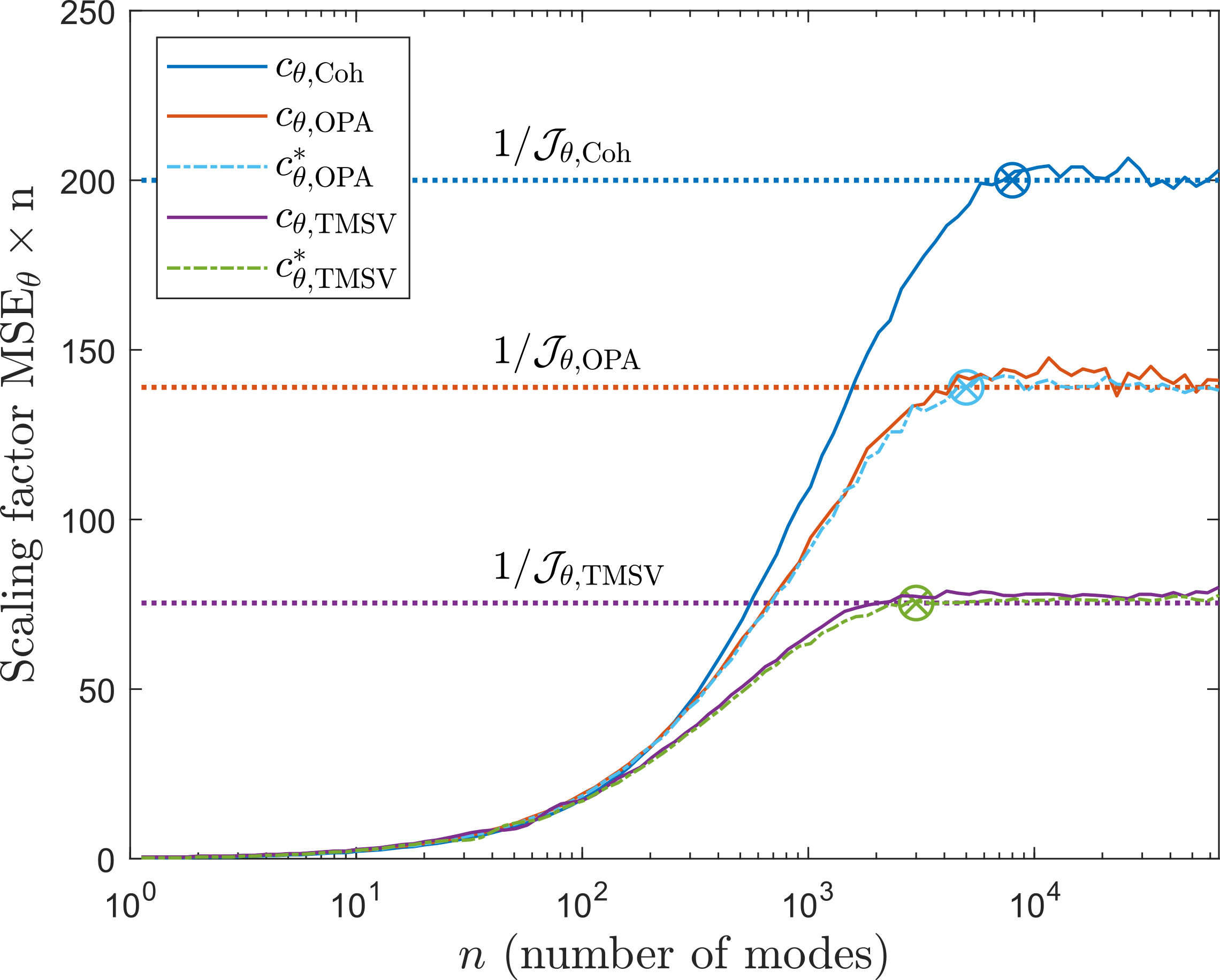}
	\caption{Scaling factor $c_\theta=n\times\text{MSE}_{\theta}$ vs.~the number of probes $n$ used to estimate $\theta$. We set $\bar{n}_{\rm S}=0.01$, $\bar{n}_{\rm B}=1$, and $\theta=0.5$. We evaluate $c_{\theta}$ at 96 logarithmically-spaced values for $n\in[1,65536]$.
The limitations of machine precision, computational run-time, and memory precluded evaluation of $c_{\theta}$ for $n>65536$.
Each data point is an average of $5\times 10^4$ results from Monte Carlo simulations.
Approximate points when scaling factors converge to the corresponding CRBs are marked with $\otimes$. 
The 95\% confidence intervals are negligibly small, and the fluctuations in the plots are due to machine-precision limits rather than stochastic variations between experiments.
	 \label{fig:simulations}}
	\vspace{-0.1cm}
\end{figure}

On the ordinate in Fig.~\ref{fig:simulations} we plot the scaling factors $c_{\theta,\text{Coh}}$, $c_{\theta,\text{OPA}}$, and $c_{\theta,\text{TMSV}}$ for the MLEs that use the output of a coherent homodyne transceiver, TMSV+OPA receiver, and TMSV+receiver derived in Section \ref{sec:optimal_receiver}, respectively.
We also plot the scaling factors $c_{\theta,\text{OPA}}^\ast$ and $c_{\theta,\text{TMSV}}^\ast$ for the corresponding receivers,  constructed with knowledge of $\theta$.
While such receivers cannot physically exist, this enables us to isolate the impact of the two-stage method on the convergence of the scaling factor.

Fig.~\ref{fig:simulations} shows that $c_{\theta,\text{TMSV}}$ converges to $\frac{1+e_{\text{two-stage}}}{\mathcal{J}_{\theta,\text{TMSV}}\left(\bar{n}_{\rm S}\right)}$ at approximately $n=3000$ modes, $c_{\theta,\text{OPA}}$ converges to $\frac{1+e_{\text{two-stage}}}{\mathcal{J}_{\theta,\text{OPA}}\left(\bar{n}_{\rm S}\right)}$ at approximately $n=5000$ modes, and $c_{\theta,\text{Coh}}$ converges to $\frac{1}{\mathcal{J}_{\theta,\text{Coh}}\left(\bar{n}_{\rm S}\right)}$ at approximately $n=8000$ modes.  The penalty from the use of the two-stage method is $e_{\text{two-stage}}\approx 0.05$, and appears to decay with $n$.
We note that $\frac{1}{\mathcal{J}_{\theta,\text{TMSV}}\left(\bar{n}_{\rm S}\right)}$ is negligibly close to the ultimate lower bound $\frac{1}{\mathcal{J}_{\theta,\text{ub}}\left(\bar{n}_{\rm S}\right)}$.

\subsection{Towards Experimental Validation}
\begin{figure}[h]
\centering
	\includegraphics[width=1\columnwidth]{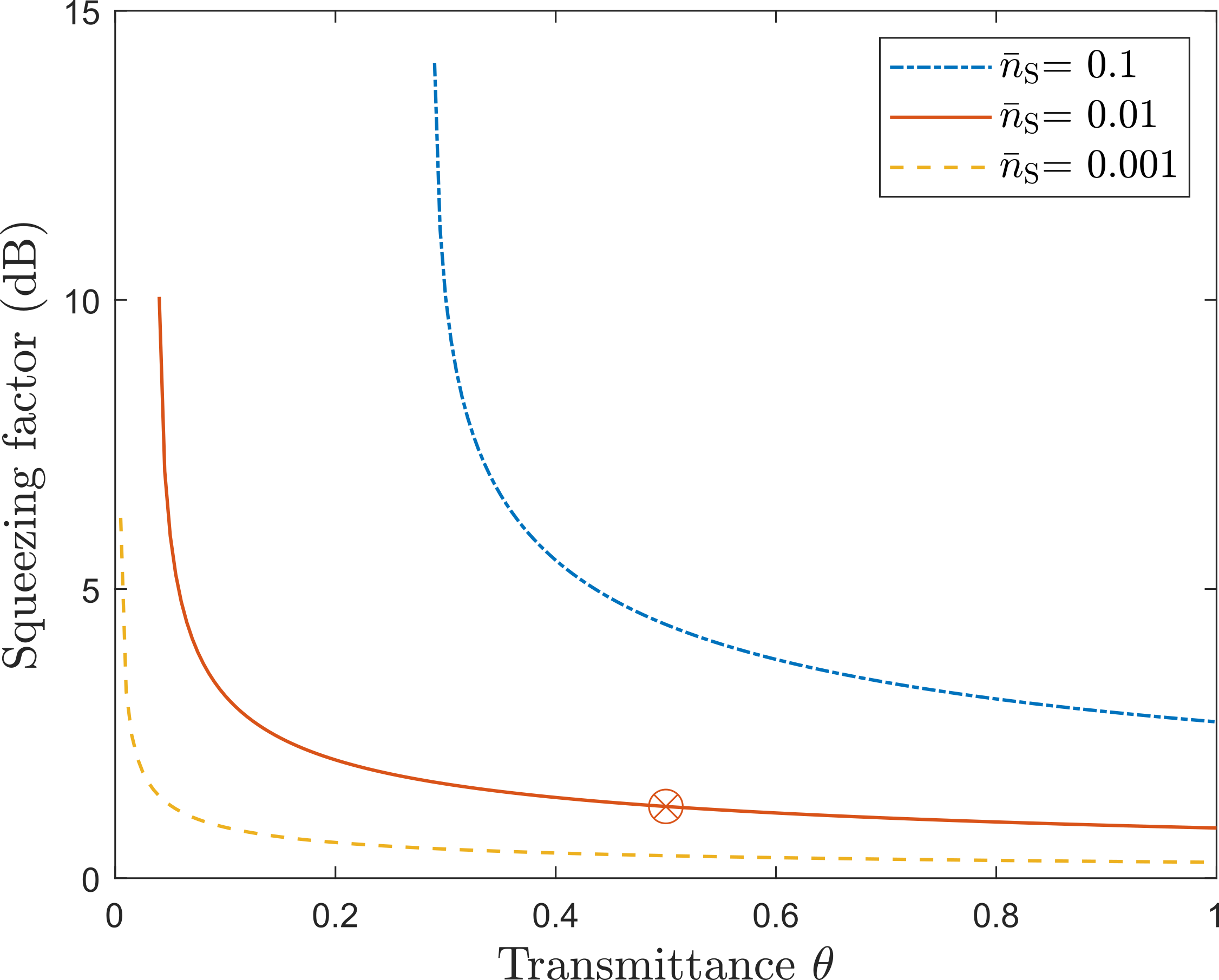}
	\caption{Squeezing factor $10\log_{10}e^{2\omega}$ vs.~transmittance $\theta$ for various values of transmitted mean photon number per mode $\bar{n}_{\rm S}$.  Thermal noise mean photon number per mode $\bar{n}_{\rm B}=1$, results for other values are very similar, as $\omega$ grows very slowly with $\bar{n}_{\rm B}$.  The squeezing factor is shown only over the regions of $\theta$ where the condition \eqref{eq:receivercondition} holds and a solution for $\omega$ exists.  These are $0.2882<\theta<1$ for $\bar{n}_{\rm S}=0.1$, $0.0385<\theta<1$ for $\bar{n}_{\rm S}=0.01$, and $0.0040<\theta<1$ for $\bar{n}_{\rm S}=0.001$. The squeezing factor required when $\bar{n}_{\rm S}=0.01$, $\bar{n}_{\rm B}=1$, and $\theta=0.5$ (Section \ref{sec:results} settings) is marked with $\otimes$.
	 \label{fig:squeezing}}
	\vspace{-0.1cm}
\end{figure}

Our results support the feasibility of near-future experimental validation of quantum-enhanced transmittance sensing.
Idler-mode storage and synchronization are technical challenges that are similar to those the previous quantum illumination experiments overcame, e.g., \cite{zhang15eesensing}.
For $\bar{n}_{\rm S}=0.01$, $\bar{n}_{\rm B}=1$, and $\theta=0.5$, the required squeezing parameter in the receiver, that we derived in Section \ref{sec:optimal_receiver}, is $\omega\approx0.1428$.
This corresponds to the squeezing factor $10\log_{10}e^{2\omega}\approx 1.24$ dB.
Fig.~\ref{fig:squeezing} further explores the squeezing factor for our receiver.
We note that $\omega$ is a very-slowly-increasing function of thermal noise photon number per mode $\bar{n}_{\rm B}$.\footnote{Numerical experiments suggest that $\omega$ grows sub-logarithmically with $\bar{n}_{\rm B}$.}
Hence, we set $\bar{n}_{\rm B}=1$.
We show that, unless transmittance $\theta$ is very small, for low transmitted mean photon number per mode $\bar{n}_{\rm S}$ (which is the regime where our sensor is optimal), substantially less than 10 dB of squeezing is needed -- a figure that has been demonstrated at 1550 nm ``telecom'' wavelength \cite{eberle13tmsv10}.
Fig.~\ref{fig:squeezing} also illustrates how the existence of $\omega$ depends on the parameters $\theta$, $\bar{n}_{\rm S}$, and $\bar{n}_{\rm B}$: when $\bar{n}_{\rm S}=0.1$ and $\bar{n}_{\rm B}=1$, the condition in \eqref{eq:receivercondition} is not satisfied for $\theta<0.2882$.
However, \eqref{eq:receivercondition} is satisfied for $\theta>0.0385$ and $\theta>0.0040$ when $\bar{n}_{\rm S}=0.01$ and $\bar{n}_{\rm S}=0.001$, respectively.
Thus, at low transmitted power, our sensor can estimate almost the entire range of $\theta$.

In our simulations, we approximate the ideal PNR by one that resolves up to 9 photons, which captures $>99\%$ of the probability mass in \eqref{eq:tmsv_opt_p}.
Such resolution has been demonstrated at 1550 nm using superconducting transition-edge sensors (TES) \cite{miller03tes}.
Fig.~\ref{fig:simulations} shows convergence to optimal MSE requiring measurement of $<10^5$ probes.
The output of a continuous-wave (cw) spontaneous parametric down-conversion (SPDC) source of entangled photons typically has an optical bandwidth of $W \sim 1$ THz \cite{kaneda19singlephotons}. 
 The quantum description of a $T$-second cw SPDC output is a tensor product of $n\approx WT$ two-mode squeezed vacuum (TMSV) states. Due to the high modes-per-second output of the SPDC, the duration of an experimental run will be governed by the electronic bandwidth of the detector, which for TES is dominated by its dead, or, reset time. The count rate of the TES reported in \cite{miller03tes} is $\approx 20 \times 10^3$ counts/s, which is far slower than commercially-available (non-photon-number-resolving) superconducting nanowire single-photon detectors (e.g., \cite{photonspot}).
However, even the TES rate allows a less-than-a-second duration for each experimental run that we simulated.

\section{Conclusion}
\label{sec:discussion}

We showed that the TMSV state asymptotically minimizes the quantum CRB for transmittance estimation in thermal noise and low-transmitted-power regime over all states (not necessarily Gaussian).
We derived a quantum CRB-achieving receiver structure for TMSV source: a two-mode squeezer followed by the PNR measurement.
Although our design is restricted to a certain range of parameters, the range corresponds to the low transmitted photon number regime where TMSV source achieves the ultimate quantum FI bound in \eqref{eq:ubQFI}.
Nevertheless, alternate receiver structures for TMSV input that are not restricted to this range of parameters should be explored.
The structure of our receiver depends on the parameter of interest, as is typical in quantum-enhanced sensing.
This necessitates a two-stage estimation approach \cite[Ch.~6.4]{hayashi17qit}, \cite{gill00twostagemeasurement,hayashi05twostagemeasurement}, however, our simulations suggest that its impact on the MSE is negligible.  

In our work, we assume thermal channel noise, availability of perfect PNR measurement, error-free idler mode storage, and its perfect synchronization with the returned probe mode.
These are also standard assumptions in theoretical investigations of quantum-enhanced sensing (e.g., \cite{monras11tmsvopt,  nair20qi, jonsson22gaussianlosssensing, lloyd08quantumillumination, tan08qigaussianstates, guha09quantumilluminationOPA, shapiro20QIstory, sanz17estimationQI}).
Strict enforcement of these assumptions in a controlled laboratory environment allowed the experimental demonstration of quantum advantage for target detection \cite{zhang15eesensing}.
Our simulation results in Section \ref{sec:simulations} suggest a high likelihood of success for a similar experimental validation of quantum advantage in transmittance sensing, with small transmitted photon number ($\bar{n}_{\rm S}\approx 0.01$), and relatively large thermal noise photon number ($\bar{n}_{\rm B}\approx 1$).
However, our future work will focus on extending both our analytic and numeric framework to account for the limitations of practical systems.
This will allow exploration of trade-offs between their complexity and performance and hasten the integration of quantum-enhanced sensing protocols into practical systems.

\appendices

\section{Proof of Theorem \ref{th:tmsv}}
\label{app:tmsv}

The proof of Theorem \ref{th:tmsv} was presented at the International Symposium on Information Theory (ISIT) 2021 and included in its proceedings \cite{gong21losssensing}.

As explained in Section \ref{sec:tmsv}, for input tensor product of TMSV states $\ket{\psi}\bra{\psi}_{IS}^{\otimes n}$, the output is a tensor-product state $\hat{\sigma}_{I^nR^n}(\theta)=\hat{\sigma}_{IR}^{\otimes n}(\theta)$ and quantum FI $\mathcal{J}_{\theta}\left(\hat{\sigma}_{I^nR^n}(\theta)\right)=n\mathcal{J}_{\theta}\left(\hat{\sigma}_{IR}(\theta)\right)$. 
Since the TMSV state and the bosonic channel are Gaussian, the output state $\hat{\sigma}_{IR}(\theta)$ is also Gaussian.
This allows the use of the symplectic formalism \cite{weedbrook12gaussianQIrmp}.
We use the $\hat{q}\hat{q}\hat{p}\hat{p}$ form for representing and evolving the covariance matrices of Gaussian states in phase-space, where $\hat{q}=\frac{\hat{a}+\hat{a}^\dagger}{\sqrt{2}}$ and $\hat{p}=\frac{\hat{a}-\hat{a}^\dagger}{\sqrt{2}}$ are the quadrature operators.
The input TMSV state's covariance matrix is:
\begin{align}
	\Sigma_{\hat{\rho}_{IS}} &=
\left[\begin{array}{cccc}
	 u_1 & u_2 & 0 & 0 \\
	u_2 & u_1 & 0 & 0 \\
	0 & 0 & u_1 & -u_2 \\
	0 & 0 & -u_2 & u_1
	\end{array}\right],
\end{align}
where $u_{1}=\bar{n}_{\rm S}+\frac{1}{2}$ and $u_2 = \sqrt{\bar{n}_{\rm S} (\bar{n}_{\rm S}+1)}$.
The action of the lossy thermal-noise bosonic channel $\mathcal{E}^{\bar{n}_{\rm T},\theta}_{S\rightarrow R}$ on the signal mode does not displace the state and results in the covariance matrix of the output:
\begin{align}
\Sigma_{\hat{\sigma}_{IR}(\theta)}&=X \Sigma_{\hat{\rho}_{IS}} X^{\rm T}+Y\\ 
&=\left[\begin{array}{cccc}
	 w_{11} & w_{12} & 0 & 0 \\
	w_{12} & w_{22} & 0 & 0 \\
	0 & 0 & w_{11} & -w_{12} \\
	0 & 0 & -w_{12} & w_{22}
\end{array}\right],
\end{align}
where $Y = \operatorname{diag}\left(0,\bar{n}_{\rm B}+\frac{1}{2}-\frac{\theta}{2},0,\bar{n}_{\rm B}+\frac{1}{2}-\frac{\theta}{2}\right)$, $X = \operatorname{diag}\left(1,\sqrt{\theta},1,\sqrt{\theta}\right)$, $A^{\rm T}$ is a transpose of $A$, and
\begin{align}
\label{eq:Appw11}	w_{11} & =\bar{n}_{\rm S}+\frac{1}{2},\\
\label{eq:Appw22}	w_{22} & = \bar{n}_{\rm B}+\theta\bar{n}_{\rm S}+\frac{1}{2}, \\
\label{eq:Appw12}	w_{12} & = \sqrt{\theta \bar{n}_{\rm S} (\bar{n}_{\rm S}+1)}.
\end{align}

We now derive the quantum FI using the method in \cite{Marian2012,Banchi2015}.  First, the Uhlmann fidelity between zero-displacement Gaussian states $\hat{\rho}_1$ and $\hat{\rho}_2$ with covariance matrices $\Sigma_1$ and $\Sigma_2$ is \cite{Marian2012}:
\begin{align}
	\label{eq:Fidelity}\mathcal{F}\left(\hat{\rho}_1,\hat{\rho}_2\right)&=\frac{1}{\sqrt{\sqrt{\Gamma}+\sqrt{\Lambda}-\sqrt{(\sqrt{\Gamma}+\sqrt{\Lambda})^2-\Delta}}},
\end{align}
where the symplectic invariants are:
\begin{align}
	\label{Delta}\Delta&=\det\left(\Sigma_1+\Sigma_2\right) \geq 1\\
	\label{Gamma}\Gamma&=16 \det\left(\Omega \Sigma_1 \Omega \Sigma_2-\frac{I_{4\times 4}}{4} \right) \geq \Delta\\
	\label{Lambda}\Lambda&=16 \det\left(\Sigma_1+\frac{\im}{2}\Omega\right) \det\left(\Sigma_2+\frac{\im}{2}\Omega\right) \geq 0,
\end{align}
$\Omega = \left[\begin{array}{cc}0_{2 \times 2} & I_{2 \times 2} \\ -I_{2 \times 2} & 0_{2 \times 2}\end{array}\right]$
is the symplectic matrix, $I_{m\times m}$ is the $m\times m$ identity matrix, and $j=\sqrt{-1}$ is the imaginary unit. The quantum FI is calculated using \eqref{eq:Fidelity} as follows:
\begin{IEEEeqnarray}{rCl}
\mathcal{J}_{\theta}\left(\hat{\sigma}_{IR}(\theta)\right)&=&
-4\left.\frac{\partial^2 \mathcal{F}\left(\hat{\sigma}_{IR}(\theta),\hat{\sigma}_{IR}(\theta+\delta_{\theta})\right)}{\partial \delta_{\theta}^2}\right|_{\delta_{\theta}=0},\label{eq:qfi_from_fidelity}
\end{IEEEeqnarray}
where \eqref{eq:qfi_from_fidelity} is derived in  \cite{Banchi2015}.
Evaluating \eqref{eq:qfi_from_fidelity} yields \eqref{eq:tmsv_qfi}.

\section{Derivation of a Quantum CRB-achieving Receiver} 
\label{app:optimal_receiver}

We obtain a quantum CRB-achieving receiver for transmittance by adapting the approach from \cite{monras07loss}.
We derive an eigendecomposion $\left\{\hat{S}(\omega)\ket{km}: k,m=0,1,\ldots\right\}$  of $\hat{\Lambda}_{\theta}$ in three steps: 1) we find an orthonormal basis $\left\{\ket{\psi_{km}}\right\}$, $k,m=0,1,\ldots$ for the output state  $\hat{\sigma}_{IR}(\theta)$; 2) we use $\left\{\ket{\psi_{km}}\right\}$ to write $\hat{\Lambda}_{\theta}$ as a linear combination of creation and annihilation operators $\hat{a}_{I}$, $\hat{a}_{R}$, $\hat{a}_{I}^\dag$, and $\hat{a}_{R}^\dag$ of the return and idler modes; and, 3) we recognize that the resulting linear combination is produced by an action of a two-mode squeezing operator $\hat{S}(\omega)$ on a number operator, yielding an expression for $\omega$.

\subsection{Orthonormal Basis for the Output State $\hat{\sigma}_{IR}(\theta)$}
\label{sec:output_orthonormal}
\newcommand{\rdel}{r}

Squeezing the two modes of $\hat{\sigma}_{IR}(\theta)$ yields $\hat{\sigma}_{IR}^{\rm d}(\theta) = \hat{S}(\zeta)  \hat{\sigma}_{IR}(\theta)\hat{S}^{\dagger}(\zeta)$ with the covariance matrix:
\begin{align}
	\Sigma_{\hat{\sigma}_{IR}^{\rm d}(\theta)}&= Z\Sigma_{\hat{\sigma}_{IR}(\theta)} Z^{\rm T}\\
	& = \left[\begin{array}{cccc}
	w^{\rm d}_{11} & w^{\rm d}_{12} & 0 & 0 \\
	w^{\rm d}_{12} & w^{\rm d}_{22} & 0 & 0 \\
	0 & 0 & w^{\rm d}_{11} & -w^{\rm d}_{12} \\
	0 & 0 & -w^{\rm d}_{12} & w^{\rm d}_{22}
	\end{array}\right],
\end{align}
where
\begin{align}
 Z& = 	\left[\begin{array}{cccc}
	\cosh(\zeta) &\sinh(\zeta) & 0 & 0 \\
	\sinh(\zeta) &\cosh(\zeta) & 0 & 0 \\
	0 & 0 &\cosh(\zeta) & -\sinh(\zeta) \\
	0 & 0 & -\sinh(\zeta) & \cosh(\zeta)
	\end{array}\right]\\
	w^{\rm d}_{11} & =w_{11}\cosh^{2}(\zeta) + w_{22}\sinh^{2}(\zeta)+w_{12} \sinh(2\zeta),\\
	w^{\rm d}_{22} & =w_{22} \cosh^{2}(\zeta) + w_{11} \sinh^{2}(\zeta) +w_{12}\sinh(2\zeta) , \\
	w^{\rm d}_{12} & = w_{12} \cosh(2\zeta) +(w_{11}+w_{22})\cosh(\zeta)\sinh(\zeta). \label{eq:wp12}
\end{align}
A value of $\zeta$ such that 
\begin{align}
\label{eq:wp120}w^{\rm d}_{12}&=0
\end{align}
makes $\hat{\sigma}_{IR}^{\rm d}(\theta)$ a thermal state that is diagonal in the Fock basis.
Note that in \eqref{eq:wp12}, $w_{11}>0$, $w_{22}>0$, $w_{12}>0$ and $\cosh(\zeta)>0$. Thus, $\sinh(\zeta)<0$. Solution of \eqref{eq:wp120} for $\zeta$ under these constraints satisfies: 
\begin{align}
\label{eq:mu}	\cosh(\zeta)&=\sqrt{\frac{1+\bar{n}_{\rm B}+\bar{n}_{\rm S}+\theta\bar{n}_{\rm S}}{2\sqrt{a}}+\frac{1}{2}}\equiv \mu,\\
\label{eq:nu}\sinh(\zeta)&=-\sqrt{\frac{1+\bar{n}_{\rm B}+\bar{n}_{\rm S}+\theta\bar{n}_{\rm S}}{2\sqrt{a}}-\frac{1}{2}}\equiv \nu.
\end{align}
This yields, in turn,
\begin{align}
\zeta&=\log\left(\sqrt{\frac{1+\bar{n}_{\rm B}+\bar{n}_{\rm S}+\theta\bar{n}_{\rm S}}{2\sqrt{a}}+\frac{1}{2}}\right.\nonumber\\
\label{eq:zeta}&\phantom{=\log\Bigg(}\left.-\sqrt{\frac{1+\bar{n}_{\rm B}+\bar{n}_{\rm S}+\theta\bar{n}_{\rm S}}{2\sqrt{a}}-\frac{1}{2}}\right),
\end{align}
where
\begin{align}
    a&=\bar{n}_{\rm B}^{2}+(1+\bar{n}_{\rm S}(1-\theta))^{2} +2 \bar{n}_{\rm B}(1+\bar{n}_{\rm S}+\theta\bar{n}_{\rm S}).
\end{align}
The covariance matrix of $\hat{\sigma}_{IR}^{\rm d}(\theta)$ can be expressed as
\begin{align}
\Sigma_{\hat{\sigma}_{IR}^{\rm d}(\theta)}&=\left[\begin{array}{cccc}
N_{1} +\frac{1}{2} & 0 & 0 & 0 \\
0 & N_{2}+\frac{1}{2} & 0 & 0 \\
0 & 0 & N_{1}+\frac{1}{2} & 0 \\
0 & 0 & 0 & N_{2}+\frac{1}{2}
\end{array}\right],
\end{align}
where the mean thermal photon numbers in each mode are:
\begin{align}
	\label{eq:N1}N_{1} &= w^{\rm d}_{11}-\frac{1}{2}=\frac{1}{2}\left (\sqrt{a}+\bar{n}_{\rm B}-1-\bar{n}_{\rm S}(1-\theta)\right )\\
\label{eq:N2}N_{2} &=w^{\rm d}_{22}-\frac{1}{2}=\frac{1}{2}\left ( \sqrt{a}-\bar{n}_{\rm B}-1+\bar{n}_{\rm S}(1-\theta)\right ).
\end{align}
Hence, using two-mode Fock basis, we have:
\begin{align}
	\hat{\sigma}_{IR}^{\rm d}(\theta) &= \sum_{km} \rdel_{km} \ket{km} \bra{km},
\end{align}
where, using the definition of $q(k,\bar{n})$ from \eqref{eq:tk},
\begin{align}
\label{eq:rkm}\rdel_{km}&\equiv q(k,N_{1})q(m,N_{2}) = \frac{N_{1}^{k}N_{2}^{m}}{(1+N_{1})^{k+1}(1+N_{2})^{m+1}}.
	\end{align}
Therefore, the output state $\hat{\sigma}_{IR}(\theta)$ is diagonal in the two-mode squeezed Fock basis $ \{ \ket{\psi_{km}} \}  $, $k,m=0,1,\ldots$:
\begin{align}
	\label{eq:output_diagonalization}\hat{\sigma}_{IR}(\theta) =& \hat{S}^{\dagger}(\zeta)  \hat{\sigma}_{IR}^{\rm d}(\theta)\hat{S}(\zeta)=\sum_{km} \rdel_{km} \ket{\psi_{km}} \bra{\psi_{km}},
\end{align}
where $ \ket{\psi_{km}} = \hat{S}^{\dagger}(\zeta) \ket{km} $ defines an orthonormal basis.

\subsection{Actions of Modal Creation and Annihilation Operators on Output State}
\label{app:master_equation}

Since the squeezing parameter $ \zeta $ is real, we have \cite[Eq.~(5.35)]{orszag16quantumotpics} 
\begin{align}
	S(\zeta)\hat{a}_{R} S^{\dagger}(\zeta) &=\mu\hat{a}_{R} +  \nu\hat{a}_{I}^{\dagger},\\
S(\zeta)\hat{a}_{R}^{\dagger} S^{\dagger}(\zeta) &= \mu\hat{a}_{R}^{\dagger} +\nu\hat{a}_{I},
\end{align}
where $ \mu$ and $\nu$ are defined in \eqref{eq:mu} and \eqref{eq:nu}, respectively.
These facts, the diagonalization of the output state $\hat{\sigma}_{IR}(\theta)$ in \eqref{eq:output_diagonalization}, and the photon-number raising and lowering properties of creation and annihilation operators allow us to derive the following six expressions for use in Appendix \ref{app:characterization}:
\begin{IEEEeqnarray}{rCl}
\IEEEeqnarraymulticol{3}{l}{\bra{\psi_{km}}2\hat{a}_{R}\hat{\sigma}_{IR}(\theta) \hat{a}_{R}^{\dagger}\ket{\psi_{st}}}\IEEEnonumber\\
&=&	\bra{km}S(\zeta)2\hat{a}_{R}S^{\dagger}(\zeta)\sum_{ul}\rdel_{ul} \ket{ul}\bra{ul}S(\zeta) \hat{a}_{R}^{\dagger}S^{\dagger}(\zeta)\ket{st}\IEEEnonumber\\
&=&\bra{km}2(\mu\hat{a}_{R} +  \nu\hat{a}_{I}^{\dagger})\sum_{ul}\rdel_{ul} \ket{ul}\bra{ul}(\mu\hat{a}_{R}^{\dagger} +\nu\hat{a}_{I})\ket{st}\IEEEnonumber\\
&=&\bra{km}2(\mu\hat{a}_{R} +  \nu\hat{a}_{I}^{\dagger})\sum_{ul}\rdel_{ul} \ket{ul}\bra{ul}(\mu\sqrt{r+1}\ket{s+1,t}\IEEEnonumber\\
&&\phantom{\bra{km}2(\mu\hat{a}_{R} +  \nu\hat{a}_{I}^{\dagger})\sum_{ul}\rdel_{ul} \ket{ul}\bra{ul}}+\nu\sqrt{t-1}\ket{s,t-1})\IEEEnonumber\\
&=&\bra{km}2(\mu\hat{a}_{R} +  \nu\hat{a}_{I}^{\dagger}) (\mu\rdel_{s+1,t}\sqrt{r+1}\ket{s+1,t}\IEEEnonumber\\
&&\phantom{\bra{km}2(\mu\hat{a}_{R} +  \nu\hat{a}_{I}^{\dagger}) (}+\nu\rdel_{s,t-1}\sqrt{t-1}\ket{s,t-1})\IEEEnonumber \\
&=&\bra{km}2(\mu\hat{a}_{R} +  \nu\hat{a}_{I}^{\dagger}) (\mu\rdel_{s+1,t}\hat{a}_{R}^{\dagger} +\nu\rdel_{s,t-1}\hat{a}_{I})\ket{st} \label{eq:Master1}\\[5pt]
\IEEEeqnarraymulticol{3}{l}{\bra{\psi_{km}}\hat{a}_{R}^{\dagger}\hat{a}_{R}\hat{\sigma}_{IR}(\theta) \ket{\psi_{st}}}\IEEEnonumber \\
&=& \bra{km}S(\zeta)\hat{a}_{R}^{\dagger}S^{\dagger}(\zeta)S(\zeta)\hat{a}_{R}S^{\dagger}(\zeta)\sum_{ul}\rdel_{ul} \ket{ul}\langle{\psi_{ul}}|{\psi_{st}}\rangle\IEEEnonumber\\
&=&\bra{km}(\mu\hat{a}_{R}^{\dagger} +\nu\hat{a}_{I})(\mu\hat{a}_{R} +  \nu\hat{a}_{I}^{\dagger})\rdel_{st}\ket{st}  \label{eq:Master2}\\[5pt]
\IEEEeqnarraymulticol{3}{l}{\bra{\psi_{km}}\hat{\sigma}_{IR}(\theta)\hat{a}_{R}^{\dagger}\hat{a}_{R} \ket{\psi_{st}}}\IEEEnonumber\\
&=& \bra{\psi_{km}}\sum_{ul}\rdel_{ul} \ket{\psi_{ul}}\bra{ul}S(\zeta)\hat{a}_{R}^{\dagger}S^{\dagger}(\zeta)S(\zeta)\hat{a}_{R}S^{\dagger}(\zeta)\ket{st}\IEEEnonumber\\
&=&\bra{km}\rdel_{km}(\mu\hat{a}_{R}^{\dagger} +\nu\hat{a}_{I})(\mu\hat{a}_{R} +  \nu\hat{a}_{I}^{\dagger})\ket{st}  \label{eq:Master3}\\[5pt]
\IEEEeqnarraymulticol{3}{l}{\bra{\psi_{km}}2\hat{a}_{R}^{\dagger}\hat{\sigma}_{IR}(\theta) \hat{a}_{R}\ket{\psi_{st}}}\IEEEnonumber\\
&= &	\bra{km}S(\zeta)2\hat{a}_{R}^{\dagger}S^{\dagger}(\zeta)\sum_{ul}\rdel_{ul} \ket{ul}\bra{ul}S(\zeta) \hat{a}_{R}S^{\dagger}(\zeta)\ket{st}\IEEEnonumber\\
&=&\bra{km}2(\mu\hat{a}_{R}^{\dagger} +\nu\hat{a}_{I})\sum_{ul}\rdel_{ul} \ket{ul}\bra{ul}(\mu\hat{a}_{R} +  \nu\hat{a}_{I}^{\dagger})\ket{st}\IEEEnonumber\\
&=&\bra{km}2(\mu\hat{a}_{R}^{\dagger} +\nu\hat{a}_{I})\sum_{ul}\rdel_{ul} \ket{ul}\bra{ul}(\mu\sqrt{k}\ket{s-1,t}\IEEEnonumber\\ 
&&\phantom{\bra{km}2(\mu\hat{a}_{R}^{\dagger} +\nu\hat{a}_{I})\sum_{ul}\rdel_{ul} \ket{ul}\bra{ul}}+\nu\sqrt{t+1}\ket{s,t+1})\IEEEnonumber\\
&=&\bra{km}2(\mu\hat{a}_{R}^{\dagger} +\nu\hat{a}_{I}) (\mu\rdel_{s-1,t}\sqrt{k}\ket{s-1,t}\IEEEnonumber\\
&&\phantom{\bra{km}2(\mu\hat{a}_{R}^{\dagger} +\nu\hat{a}_{I}) (}+\nu\rdel_{s,t+1}\sqrt{t+1}\ket{s,t+1})\IEEEnonumber \\
&=&\bra{km}2(\mu\hat{a}_{R}^{\dagger} +\nu\hat{a}_{I}) (\mu\rdel_{s-1,t}\hat{a}_{R} +\nu\rdel_{s,t+1}\hat{a}_{I}^{\dagger})\ket{st} \label{eq:Master4}\\[5pt]
\IEEEeqnarraymulticol{3}{l}{\bra{\psi_{km}}\hat{a}_{R}\hat{a}_{R}^{\dagger}\hat{\sigma}_{IR}(\theta) \ket{\psi_{st}}}\IEEEnonumber \\
&=& \bra{km}S(\zeta)\hat{a}_{R}S^{\dagger}(\zeta)S(\zeta)\hat{a}_{R}^{\dagger}S^{\dagger}(\zeta)\sum_{ul}\rdel_{ul} \ket{ul}\langle{\psi_{ul}}|{\psi_{st}}\rangle\IEEEnonumber\\
&=&\bra{km}(\mu\hat{a}_{R} +  \nu\hat{a}_{I}^{\dagger})(\mu\hat{a}_{R}^{\dagger} +\nu\hat{a}_{I})\rdel_{st}\ket{st}  \label{eq:Master5}\\[5pt]
\IEEEeqnarraymulticol{3}{l}{\bra{\psi_{km}}\hat{\sigma}_{IR}(\theta)\hat{a}_{R}\hat{a}_{R}^{\dagger} \ket{\psi_{st}}}\IEEEnonumber \\
&=& \bra{\psi_{km}}\sum_{ul}\rdel_{ul} \ket{\psi_{ul}}\bra{ul}S(\zeta)\hat{a}_{R}S^{\dagger}(\zeta)S(\zeta)\hat{a}_{R}^{\dagger}S^{\dagger}(\zeta)\ket{st}\IEEEnonumber\\
&=&\bra{km}\rdel_{km}(\mu\hat{a}_{R} +  \nu\hat{a}_{I}^{\dagger})(\mu\hat{a}_{R}^{\dagger} +\nu\hat{a}_{I})\ket{st}  \label{eq:Master6}
\end{IEEEeqnarray}

\subsection{Characterization of SLD $\hat{\Lambda}_{\theta}$}
\label{app:characterization}

First, we use \eqref{eq:sld_def} to relate the $ kmst $-th term of the SLD operator $\hat{\Lambda}_{\theta}$ in the basis $ \{ \ket{\psi_{km}} \} $ to the corresponding term of the derivative $\frac{d \hat{\sigma}_{IR}(\theta) }{d \theta}$:
\begin{align}
\bra{\psi_{km}} \frac{d \hat{\sigma}_{IR}(\theta) }{d \theta} \ket{\psi_{st}}&= \frac{1}{2}\bra{\psi_{km}} \hat{\sigma}_{IR}(\theta)	\hat{\Lambda}_{\theta}\ket{\psi_{st}}\nonumber\\ 
&\phantom{=}+ 	\frac{1}{2}\bra{\psi_{km}}\hat{\Lambda}_{\theta}\hat{\sigma}_{IR}(\theta)\ket{\psi_{st}}\\
	&= \frac{\rdel_{km}+\rdel_{st}}{2}  \bra{\psi_{km}}	\hat{\Lambda}_{\theta} \ket{\psi_{st}}. \label{eq:deta}
\end{align}
Thus, the SLD operator is expressed as follows:
\begin{align}
		\hat{\Lambda}_{\theta} 
	& =  \sum_{kmst} \bra{\psi_{km}}  \frac{2\frac{d \hat{\sigma}_{IR}(\theta) }{d \theta} }{\rdel_{km}+\rdel_{st}} \ket{\psi_{st}} \ket{\psi_{km}}  \bra{\psi_{st}}.\label{eq:SLD1}
\end{align}

The probe state evolving in a thermal bath is characterized by the Lindblad Master equation \cite[Ch.~4]{ferraro05gaussian}:
\begin{align}
\frac{d \hat{\sigma}_{IR}(\theta) }{d t} = \frac{\gamma}{2}\left [ (\bar{n}_{\rm B}+1) \hat{\mathcal{L}}_{R}[\hat{a}]+\bar{n}_{\rm B}\hat{\mathcal{L}}_{R}[\hat{a}^{\dagger}]\right ] \hat{\sigma}_{IR}(\theta), \label{eq:dt}
\end{align}
where the superoperator $\hat{\mathcal{L}}_{R}[\cdot]$ is defined as follows:
\begin{align}
\hat{\mathcal{L}}_{R}[\hat{a}] \hat{\sigma}_{IR}(\theta) & = 2 \hat{a}_{R}\hat{\sigma}_{IR}(\theta) \hat{a}_{R}^{\dagger} - \hat{a}_{R}^{\dagger}\hat{a}_{R}\hat{\sigma}_{IR}(\theta)\nonumber\\
 &\phantom{=} - \hat{\sigma}_{IR}(\theta)\hat{a}_{R}^{\dagger}\hat{a}_{R}\\
\hat{\mathcal{L}}_{R}[\hat{a}^{\dagger}] \hat{\sigma}_{IR}(\theta) & = 2 \hat{a}_{R}^{\dagger}\hat{\sigma}_{IR} (\theta)\hat{a}_{R} - \hat{a}_{R}\hat{a}_{R}^{\dagger}\hat{\sigma}_{IR}(\theta) \nonumber\\
 &\phantom{=}- \hat{\sigma}_{IR}(\theta)\hat{a}_{R}\hat{a}_{R}^{\dagger}.
\end{align}
The dissipation rate $\gamma$ satisfies $\exp\left (-\frac{\gamma}{2}t\right ) = \sqrt\theta$, which, in turn, implies:
\begin{align}
\frac{dt}{d\theta} = \frac{1}{\gamma \theta}. \label{eq:dtdeta}
\end{align}

Employing the chain rule in \eqref{eq:dt} using \eqref{eq:dtdeta}, and substituting the result into \eqref{eq:SLD1} yields:

\begin{align}
	\hat{\Lambda}_{\theta}	&=  \sum_{kmst} \bra{\psi_{km}}  \frac{2\frac{d \hat{\sigma}_{IR} (\theta)}{d t} \frac{dt}{d\theta} }{\rdel_{km}+\rdel_{st}} \ket{\psi_{st}} \ket{\psi_{km}}  \bra{\psi_{st}}\\
	  &=\frac{(\bar{n}_{\rm B}+1)}{\theta} \sum_{kmst} \frac{\bra{\psi_{km}} \hat{\mathcal{L}}_{R}[\hat{a}]\hat{\sigma}_{IR}(\theta)\ket{\psi_{st}}  }{\rdel_{km}+\rdel_{st}}\ket{\psi_{km}}\bra{\psi_{st}}\nonumber\\
\label{eq:SLD_Master_eq}	 &\phantom{=}+\frac{\bar{n}_{\rm B}}{\theta} \sum_{kmst} \frac{\bra{\psi_{km}} \hat{\mathcal{L}}_{R}[\hat{a}^{\dagger}]\hat{\sigma}_{IR}(\theta)\ket{\psi_{st}}  }{\rdel_{km}+\rdel_{st}}\ket{\psi_{km}}\bra{\psi_{st}}
\end{align}

We analyze the two summations in \eqref{eq:SLD_Master_eq} separately.  First,
\begin{IEEEeqnarray}{rCl}
\IEEEeqnarraymulticol{3}{l}{\frac{\bra{\psi_{km}} \hat{\mathcal{L}}_{R}[\hat{a}]\hat{\sigma}_{IR}(\theta)\ket{\psi_{st}}  }{\rdel_{km}+\rdel_{st}}}\IEEEnonumber\\
 &=& \frac{\bra{\psi_{km}} 2 \hat{a}_{R}\hat{\sigma}_{IR}(\theta)\hat{a}_{R}^{\dagger}\ket{\psi_{st}}}{\rdel_{km}+\rdel_{st}}\IEEEnonumber\\ 
&\phantom{=}&-\frac{\bra{\psi_{km}}\hat{a}_{R}^{\dagger}\hat{a}_{R}\hat{\sigma}_{IR} (\theta)\ket{\psi_{st}}}{\rdel_{km}+\rdel_{st}}\IEEEnonumber\\
&\phantom{=}&- \frac{\bra{\psi_{km}}\hat{\sigma}_{IR}(\theta)\hat{a}_{R}^{\dagger}\hat{a}_{R}\ket{\psi_{st}}  }{\rdel_{km}+\rdel_{st}}\\
&=&  \frac{\bra{km}2(\mu\hat{a}_{R} +  \nu\hat{a}_{I}^{\dagger}) (\mu\rdel_{s+1,t}\hat{a}_{R}^{\dagger} +\nu\rdel_{s,t-1}\hat{a}_{I})\ket{st}}{\rdel_{km}+\rdel_{st}}\IEEEnonumber\\
&&- \bra{km}(\mu\hat{a}_{R}^{\dagger} +\nu\hat{a}_{I})(\mu\hat{a}_{R} +  \nu\hat{a}_{I}^{\dagger})\ket{st} \label{eq:first,hatK}\\
&=& \bra{km}\left (\frac{2\rdel_{s+1,t}}{\rdel_{km}+\rdel_{st}}-1\right)\mu^2\hat{a}_{R}^{\dagger}\hat{a}_{R}\ket{st}\IEEEnonumber\\
&\phantom{=}&+ \bra{km}\left(\frac{2\rdel_{s+1,t}}{\rdel_{km}+\rdel_{st}}-1\right)\nu\mu\hat{a}_{I}^{\dagger}\hat{a}_{R}^{\dagger}\ket{st}\IEEEnonumber\\
&\phantom{=}&+\bra{km}\left(\frac{2\rdel_{s,t-1}}{\rdel_{km}+\rdel_{st}}-1\right)\mu\nu\hat{a}_{R} \hat{a}_{I}\ket{st} \IEEEnonumber\\
&\phantom{=}&+\bra{km}\left(\frac{2\rdel_{s,t-1}}{\rdel_{km}+\rdel_{st}}-1\right)\nu^2\hat{a}_{I}^{\dagger}\hat{a}_{I}\ket{st}\IEEEnonumber\\
&\phantom{=}&+\left(\frac{2\mu^2\rdel_{s+1,t}}{\rdel_{km}+\rdel_{st}}-\nu^2\right)\left\langle km\middle| st\right\rangle,\label{eq:second,hatK}
\end{IEEEeqnarray}
where $\mu$ an $\nu$ are defined in \eqref{eq:mu} and \eqref{eq:nu}, respectively, \eqref{eq:first,hatK} is derived using \eqref{eq:Master1}-\eqref{eq:Master6} in Appendix \ref{app:master_equation}, and \eqref{eq:second,hatK} is due to the commutation relation $\left[\hat{a},\hat{a}^{\dagger}\right]=\hat{I}$ (with $\hat{I}$ denoting the identity operator), and rearrangement of terms. Observe that the first, fourth, and fifth terms in $ \eqref{eq:second,hatK} $ are not zero only when $\{k = s, m=t\}$, while the second and third terms are not zero when $\{k= s+1, m=t+1\}$ and $\{k = s-1, m=t-1\}$, respectively.
Since
\begin{align}
\left.\frac{2\rdel_{s+1,t}}{\rdel_{km}+\rdel_{st}}\right|_{k=s,m=t} -1&= -\frac{1}{1+N_{1}}\\
\left.\frac{2\rdel_{s+1,t}}{\rdel_{km}+\rdel_{st}}\right|_{k=s+1,m=t+1}-1&=\frac{N_{1}-N_{2}-1}{2N_{1}N_{2}+N_{1}+N_{2}+1}\\
\left.\frac{2\rdel_{s,t-1}}{\rdel_{km}+\rdel_{st}}\right|_{k=s-1,m=t-1}-1&=\frac{N_{1}-N_{2}-1}{2N_{1}N_{2}+N_{1}+N_{2}+1}\\
\left.\frac{2\rdel_{s,t-1}}{\rdel_{km}+\rdel_{st}}\right|_{k=s,m=t}-1&= \frac{1}{N_{2}},
\end{align}
with $N_{1}$ and $N_{2}$ defined in \eqref{eq:N1} and \eqref{eq:N2}, respectively, we have:
\begin{multline}
\sum_{kmst} \frac{\bra{\psi_{km}} \hat{\mathcal{L}}_{R}[\hat{a}]\hat{\sigma}_{IR}(\theta)\ket{\psi_{st}}  }{\rdel_{km}+\rdel_{st}}\ket{\psi_{km}}\bra{\psi_{st}} \\ = \hat{S}^{\dagger}(\zeta)  \hat{K}_1   \hat{S}(\zeta),\label{eq:hatk1}
\end{multline}
where
\begin{align}
    \hat{K}_1 &=  \left(\frac{N_{1}-N_{2}-1}{2N_{1}N_{2}+N_{1}+N_{2}+1}\right)\nu\mu\left(\hat{a}_{I}^{\dagger}\hat{a}_{R}^{\dagger}+\hat{a}_{R} \hat{a}_{I}\right)\nonumber\\ 
    &\phantom{=}-\frac{\mu^2\hat{a}_{R}^{\dagger}\hat{a}_{R}}{1+N_{1}}+\frac{\nu^2}{N_{2}}\hat{a}_{I}^{\dagger}\hat{a}_{I}+\left(\mu^2\frac{N_{1}}{1+N_{1}}-\nu^2\right)\hat{I}.
\end{align}
Now, the $kmst$-th term in the second summation in \eqref{eq:SLD_Master_eq} is:
\begin{IEEEeqnarray}{rCl}
\IEEEeqnarraymulticol{3}{l}{\frac{\bra{\psi_{km}} \hat{\mathcal{L}}_{R}[\hat{a}^{\dagger}]\hat{\sigma}_{IR}(\theta)\ket{\psi_{st}}  }{\rdel_{km}+\rdel_{st}}}\IEEEnonumber\\
 &=& \frac{\bra{\psi_{km}} 2 \hat{a}_{R}^{\dagger}\hat{\sigma}_{IR}(\theta)\hat{a}_{R}\ket{\psi_{st}}}{\rdel_{km}+\rdel_{st}}\IEEEnonumber\\ 
&\phantom{=}&-\frac{\bra{\psi_{km}}\hat{a}_{R}\hat{a}_{R}^{\dagger}\hat{\sigma}_{IR} (\theta)\ket{\psi_{st}}}{\rdel_{km}+\rdel_{st}}\IEEEnonumber\\
&\phantom{=}&- \frac{\bra{\psi_{km}}\hat{\sigma}_{IR}(\theta)\hat{a}_{R}\hat{a}_{R}^{\dagger}\ket{\psi_{st}}  }{\rdel_{km}+\rdel_{st}}\\
&=&  \frac{\bra{km}2(\mu\hat{a}_{R}^{\dagger} +  \nu\hat{a}_{I}) (\mu\rdel_{s-1,t}\hat{a}_{R} +\nu\rdel_{s,t+1}\hat{a}_{I}^{\dagger})\ket{st}}{\rdel_{km}+\rdel_{st}}\IEEEnonumber\\
&&- \bra{km}(\mu\hat{a}_{R} +\nu\hat{a}_{I}^{\dagger})(\mu\hat{a}_{R}^{\dagger} +  \nu\hat{a}_{I})\ket{st} \label{eq:third,hatK}\\
&=& \bra{km}\left (\frac{2\rdel_{s-1,t}}{\rdel_{km}+\rdel_{st}}-1\right)\mu^2\hat{a}_{R}^{\dagger}\hat{a}_{R}\ket{st}\IEEEnonumber\\
&\phantom{=}&+ \bra{km}\left(\frac{2\rdel_{s-1,t}}{\rdel_{km}+\rdel_{st}}-1\right)\nu\mu\hat{a}_{I}\hat{a}_{R}\ket{st}\IEEEnonumber\\
&\phantom{=}&+\bra{km}\left(\frac{2\rdel_{s,t+1}}{\rdel_{km}+\rdel_{st}}-1\right)\mu\nu\hat{a}_{R}^{\dagger} \hat{a}_{I}^{\dagger}\ket{st} \IEEEnonumber\\
&\phantom{=}&+\bra{km}\left(\frac{2\rdel_{s,t+1}}{\rdel_{km}+\rdel_{st}}-1\right)\nu^2\hat{a}_{I}^{\dagger}\hat{a}_{I}\ket{st}\IEEEnonumber\\
&\phantom{=}&+\left(\frac{2\nu^2\rdel_{s+1,t}}{\rdel_{km}+\rdel_{st}}-\mu^2\right)\left\langle km\middle| st\right\rangle,\label{eq:fourth,hatK}
\end{IEEEeqnarray}
where $\mu$ an $\nu$ are defined in \eqref{eq:mu} and \eqref{eq:nu}, respectively, \eqref{eq:third,hatK} is derived using \eqref{eq:Master1}-\eqref{eq:Master6} in Appendix \ref{app:master_equation}, and \eqref{eq:fourth,hatK} is due to the commutation relation $\left[\hat{a},\hat{a}^{\dagger}\right]=\hat{I}$, and rearrangement of terms. Observe that the first, fourth, and fifth terms in $ \eqref{eq:fourth,hatK} $ are not zero only when $\{k = s, m=t\}$, while the second and third terms are not zero when $\{k= s-1, m=t-1\}$ and $\{k = s+1, m=t+1\}$, respectively.
Since
\begin{align}
\left.\frac{2\rdel_{s-1,t}}{\rdel_{km}+\rdel_{st}}\right|_{k=s,m=t} -1&= \frac{1}{N_{1}}\\
\left.\frac{2\rdel_{s-1,t}}{\rdel_{km}+\rdel_{st}}\right|_{k=s-1,m=t-1}-1&=\frac{N_{2}-N_{1}-1}{2N_{1}N_{2}+N_{1}+N_{2}+1}\\
\left.\frac{2\rdel_{s,t+1}}{\rdel_{km}+\rdel_{st}}\right|_{k=s+1,m=t+1}-1&=\frac{N_{2}-N_{1}-1}{2N_{1}N_{2}+N_{1}+N_{2}+1}\\
\left.\frac{2\rdel_{s,t+1}}{\rdel_{km}+\rdel_{st}}\right|_{k=s,m=t}-1&= -\frac{1}{1+N_{2}},
\end{align}
we have:
\begin{multline}
\sum_{kmst} \frac{\bra{\psi_{km}} \hat{\mathcal{L}}_{R}[\hat{a}^{\dagger}]\hat{\sigma}_{IR}(\theta)\ket{\psi_{st}}  }{\rdel_{km}+\rdel_{st}}\ket{\psi_{km}}\bra{\psi_{st}}\\ = \hat{S}^{\dagger}(\zeta)  \hat{K}_2 \hat{S}(\zeta),\label{eq:hatk2}
\end{multline}
where
\begin{align}
    \hat{K}_2 &= \left(\frac{N_{2}-N_{1}-1}{2N_{1}N_{2}+N_{1}+N_{2}+1}\right)\nu\mu\left(\hat{a}_{I}^{\dagger}\hat{a}_{R}^{\dagger}+\hat{a}_{R} \hat{a}_{I}\right)\nonumber\\ 
    &\phantom{=}+\frac{\mu^2\hat{a}_{R}^{\dagger}\hat{a}_{R}}{N_{1}}-\frac{\nu^2}{1+N_{2}}\hat{a}_{I}^{\dagger}\hat{a}_{I}+\left(\nu^2\frac{N_{2}}{1+N_{2}}-\mu^2\right)\hat{I}.
\end{align}
Combining \eqref{eq:hatk1} and \eqref{eq:hatk2} yields:
\begin{align}
		\hat{\Lambda}_{\theta} &= \frac{1}{\theta} \hat{S}^{\dagger}(\zeta)\hat{K}\hat{S}(\zeta), \label{eq:SLD2}
\end{align}
where
\begin{align}
\hat{K}	&=(\bar{n}_{\rm B}+1)\hat{K}_1 + \bar{n}_{\rm B} \hat{K}_2\\
&=  C \hat{a}_{R}^{\dagger}\hat{a}_{R} +D \hat{a}_{I}^{\dagger}\hat{a}_{I} + E (\hat{a}_{I}^{\dagger}\hat{a}_{R}^{\dagger}+\hat{a}_{R} \hat{a}_{I}) +F\hat{I},
\end{align}
and real scalars
\begin{align}
\label{eq:C}	C&=\mu^2 \frac{\bar{n}_{\rm B} -N_{1}}{N_{1}(1+N_{1})}\\
\label{eq:D}	D&=\nu^2 \left ( \frac{\bar{n}_{\rm B}+1+N_{2}}{N_{2}(1+N_{2})} \right )\\
\label{eq:E} E&= \mu\nu\left(\frac{N_{1}-N_{2}-2\bar{n}_{\rm B} - 1}{2N_{1}N_{2}+N_{1}+N_{2}+1} \right)\\
\label{eq:F} F&= \mu^2 \left(\frac{N_{1}-\bar{n}_{\rm B}}{1+N_{1}}\right) -\nu^2 \left(\frac{\bar{n}_{\rm B}+1+N_{2}}{1+N_{2}}\right),
\end{align}
with $\mu$, $\nu$, $N_1$, and $N_2$ defined in \eqref{eq:mu}, \eqref{eq:nu}, \eqref{eq:N1}, and \eqref{eq:N2}, respectively.

\subsection{Eigenbasis of the SLD $\hat{\Lambda}_{\theta}$}
\label{app:eigenbasis}
Application of a two-mode squeezing operator $\hat{S}\left(\lambda e^{\im \theta_{\lambda}}\right)$ to a photon number operator $\hat{a}_{I}^{\dagger}\hat{a}_{I}$ results in:
\begin{IEEEeqnarray}{rCl}
\IEEEeqnarraymulticol{3}{l}{\hat{S}\left(\lambda e^{\im \theta_{\lambda}}\right)\hat{a}_{R}^{\dagger}\hat{a}_{R}\hat{S}^{\dagger}\left(\lambda e^{\im \theta_{\lambda}}\right)}\IEEEnonumber\\
&=&\hat{S}\left(\lambda e^{\im \theta_{\lambda}}\right)\hat{a}_{R}^{\dagger}\hat{S}^{\dagger}\left(\lambda e^{\im \theta_{\lambda}}\right)\hat{S}\left(\lambda e^{\im \theta_{\lambda}}\right)\hat{a}_{R}\hat{S}^{\dagger}\left(\lambda e^{\im \theta_{\lambda}}\right)\IEEEnonumber\\
	&=&  (\kappa \hat{a}_{R}^{\dagger} +  \xi e^{-\im \theta_{\lambda}}\hat{a}_{I})(\kappa \hat{a}_{R} +  \xi e^{\im \theta_{\lambda}} \hat{a}_{I}^{\dagger})\IEEEnonumber\\
	&=& \kappa^2 \hat{a}_{R}^{\dagger }\hat{a}_{R} +\kappa \xi (e^{\im \theta_{\lambda}}\hat{a}_{R}\hat{a}_{I}+e^{-\im \theta_{\lambda}}\hat{a}_{R}^{\dagger}\hat{a}_{I}^{\dagger} )+\xi^2 \hat{a}_{I}\hat{a}_{I}^{\dagger },\qquad\label{eq:s1}
\end{IEEEeqnarray}
where $\kappa = \cosh \lambda$ and $\xi = \sinh \lambda$.
Similarly,
\begin{IEEEeqnarray}{rCl}
\IEEEeqnarraymulticol{3}{l}{\hat{S}\left(\lambda e^{\im \theta_{\lambda}}\right)\hat{a}_{I}^{\dagger}\hat{a}_{I}\hat{S}^{\dagger}\left(\lambda e^{\im \theta_{\lambda}}\right)}\IEEEnonumber\\
&=& \kappa^2 \hat{a}_{I}^{\dagger } \hat{a}_{I}+\kappa \xi (e^{-\im \theta_{\lambda}}\hat{a}_{I}\hat{a}_{R}+e^{\im \theta_{\lambda}}\hat{a}_{I}^{\dagger}\hat{a}_{R}^{\dagger} ) +\xi^2 \hat{a}_{R}\hat{a}_{R}^{\dagger }.\qquad \label{eq:s2}
\end{IEEEeqnarray}
Thus, provided scalars $\lambda$, $F^{\prime}$, $T_1$, $T_2$ exist, we can write:
\begin{align}
	\hat{K} &= F^{\prime}\hat{I} + \hat{S}(\lambda)(T_1\hat{a}_{R}^{\dagger}\hat{a}_{R}+T_2\hat{a}_{I}^{\dagger}\hat{a}_{I})\hat{S}^{\dagger}(\lambda)\label{eq:Khat_from_numberop}\\
		&= F^{\prime}\hat{I} + (T_1\kappa \xi+ T_2\kappa \xi)(\hat{a}_{R}\hat{a}_{I}+\hat{a}_{R}^{\dagger}\hat{a}_{I}^{\dagger} )+T_1\xi^{2} + T_2\xi^2\nonumber\\
	&\phantom{=}+ (T_1\kappa^2 + T_2 \xi^2) \hat{a}_{R}^{\dagger }\hat{a}_{R}+ (T_1\xi^2+ T_2 \kappa^2)  \hat{a}_{I}^{\dagger }\hat{a}_{I},  \label{eq:LHSu}
\end{align}
where \eqref{eq:LHSu} is from substituting \eqref{eq:s1} and \eqref{eq:s2} in \eqref{eq:Khat_from_numberop} and rearranging terms.
Note that it is necessary that $\theta_{\lambda}=0$, which means that scalars $\lambda$, $F^{\prime}$, $T_1$, $T_2$ must be real and satisfy:
\begin{align}
	C& = T_1\kappa^2 + T_2 \xi^2\label{eq:vc}\\
	D& = T_1\xi^2+ T_2 \kappa^2\label{eq:vd}\\
E& =(T_1+ T_2)\kappa \xi = (T_1+ T_2) \frac{\sinh 2\lambda}{2}, \label{eq:ve}\\
	F& = F^{\prime} + T_1\xi^{2} + T_2\xi^2
\end{align}
where scalars $C$, $D$, $E$, and $F$ are given in \eqref{eq:C}-\eqref{eq:F}.
Now,
\begin{align}
	C - D &= T_1(\kappa^2 - \xi^2) + T_2(\xi^2 - \kappa^2)= T_1 - T_2, \label{eq:C - D}\\
		C + D &= T_1(\kappa^2 + \xi^2) + T_2(\xi^2 + \kappa^2)= (T_1+T_2)\cosh 2 \lambda, \label{eq:C + D}
\end{align}
Furthermore,
\begin{align}
(C+D)^2 - 4 E^2&=(T_1+T_2)^2 (\cosh^2 2\lambda - \sinh^2 2\lambda)\nonumber\\
&=(T_1+T_2)^2.\label{eq:t1 + t2}
\end{align}
Using \eqref{eq:C - D}-\eqref{eq:t1 + t2}, we obtain:
\begin{align}
\label{eq:tau1}T_1 &= \frac{1}{2} (\sqrt{(C+D)^{2}-4 E^2} + C-D)\\
\label{eq:tau2}T_2 &= \frac{1}{2} (\sqrt{(C+D)^{2}-4 E^2} - C+D)\\
\label{eq:cosh2lambda}\cosh 2\lambda &= \frac{C + D}{\sqrt{(C+D)^2-4 E^2}}\\
\label{eq:sinh2lambda}\sinh 2\lambda &= \frac{2E}{\sqrt{(C+D)^2-4 E^2}}\\
\label{eq:Fprime}F^{\prime} &= F -(T_1+T_2)\sinh^{2}(\lambda).
\end{align}
Finally, we can show that
\begin{IEEEeqnarray}{rCl}
\IEEEeqnarraymulticol{3}{l}{\hat{\Lambda}_{\theta} S^{\dagger}(\zeta) S(\lambda)\ket{km}}\IEEEnonumber\\
&=&\frac{1}{\theta} S^{\dagger}(\zeta)  \hat{K}   S(\zeta) S^{\dagger}(\zeta) S(\lambda)\ket{km} \IEEEnonumber \\
&=& \frac{1}{\theta} S^{\dagger}(\zeta) [ F^{\prime} + S(\lambda) (T_1\hat{a}_{R}^{\dagger}\hat{a}_{R}+T_2\hat{a}_{I}^{\dagger}\hat{a}_{I}) ^{\dagger}S(\lambda)^{\dagger}] S(\lambda)\ket{km}\IEEEnonumber\\
&=&\frac{1}{\theta}\left ( F^{\prime} + T_1 k+ T_2m \right )  S^{\dagger}(\zeta) S(\lambda)\ket{km},
\end{IEEEeqnarray}
where
\begin{align}
\label{eq:lambda}\lambda&=\frac{1}{2}\log\left(\frac{2E+|C+D|}{\sqrt{(C+D)^2-4 E^2}}\right).
\end{align}
Thus, $\hat{S}(\omega)\ket{km}$, $\omega=\lambda - \zeta$ is an eigenvector of the SLD $\hat{\Lambda}_{\theta}$.

\section{Fock-basis Representation of Two-mode Squeezer}
\label{app:focksqueezing}

To derive the representation in \eqref{eq:focksqueezing} of the two-mode squeezing operator $\hat{S}(\omega_1)$ in the two-mode Fock (photon number) basis $\left\{\ket{km}:k,m=0,1,\ldots\right\}$, note that \cite[Eq.~(1.233)]{kok2010introduction}:
\begin{align}
    S(\omega_1) &= e^{-\omega_1 \hat{a}_R^\dagger \hat{a}_I^\dagger+ \omega_1^* \hat{a}_R\hat{a}_I}\\
    &= e^{-\tau_1 \hat{a}_R^\dagger\hat{a}_I^\dagger} \nu_1^{-\hat{a}_R^\dagger\hat{a}_R - \hat{a}_I^\dagger\hat{a}_I-\hat{I}} e^{\tau_1^*\hat{a}_R\hat{a}_I},
\end{align}
where $\hat{I}$ denotes identity operator, and
\begin{align}
    \tau_1 &= \frac{\omega_1}{|\omega_1|} \tanh |\omega_1|\\
    \nu_1 &= \cosh|\omega_1|.
\end{align}
Now,
\begin{IEEEeqnarray}{rCl}
	\IEEEeqnarraymulticol{3}{l}{
    S(\omega_1)\ket{km}}\IEEEnonumber \\
    &=& e^{-\tau_1 \hat{a}_R^\dagger\hat{a}_I^\dagger} \nu_1^{-\hat{a}_R^\dagger\hat{a}_R - \hat{a}_I^\dagger\hat{a}_I-\hat{I}} e^{\tau_1^*\hat{a}_R\hat{a}_I}\ket{km} \\
    & =& e^{-\tau_1 \hat{a}_R^\dagger\hat{a}_I^\dagger} \nu_1^{-\hat{a}_R^\dagger\hat{a}_R - \hat{a}_I^\dagger\hat{a}_I-\hat{I}}\IEEEnonumber\\
    &&\times \sum_{u=0}^{\infty} \frac{(\tau_1^*)^u}{u!} (\hat{a}_R\hat{a}_I)^u \ket{km} \label{eq:TMS_NS_t1}\\
    &=& \sum_{u=0}^{\min(k,m)} e^{-\tau_1 \hat{a}_R^\dagger\hat{a}_I^\dagger} \nu_1^{-\hat{a}_R^\dagger\hat{a}_R - \hat{a}_I^\dagger\hat{a}_I-\hat{I}}  \frac{(\tau_1^*)^u}{u!} \IEEEnonumber \\ && \times \sqrt{\frac{k!m!}{(k-u)!(m-u)!}}\ket{k-u,m-u}\label{eq:TMS_NS_m0}\\
    &=& \sum_{u=0}^{\min(k,m)} e^{-\tau_1 \hat{a}_R^\dagger\hat{a}_I^\dagger} \nu_1^{-k - m + 2u - 1}  \frac{(\tau_1^*)^u}{u!} \IEEEnonumber \\ && \times \sqrt{\frac{k!m!}{(k-u)!(m-u)!}}\ket{k-u,m-u}\label{eq:TMS_NS_m1}\\
    & =& \sum_{u=0}^{\min(m,n)} \sum_{l=0}^{\infty} \frac{(-\tau_1)^l}{l!}  (\hat{a}_R^\dagger \hat{a}_I^\dagger)^l \nu_1^{-k - m + 2u - 1}  \IEEEnonumber \\ && \times \frac{(\tau_1^*)^u}{u!} \sqrt{\frac{k!m!}{(k-u)!(m-u)!}}\ket{k-u,m-u} \label{eq:TMS_NS_t2}\\
    &=& \sum_{u=0}^{\min(k,m)} \sum_{l=0}^{\infty} \frac{(-\tau_1)^l}{l!}   \nu_1^{-k - m + 2u - 1}  \frac{(\tau_1^*)^u}{u!}  \IEEEnonumber \\ &&\phantom{=} \times
    \sqrt{\frac{(k-u+l)!(m-u+l)!}{(k-u)!(m-u)!}}   \IEEEnonumber \\ &&\phantom{=} \times \sqrt{\frac{k!m!}{(k-u)!(m-u)!}}\IEEEnonumber \\ &&\phantom{=} \times\ket{k-u+l,m-u+l}, \label{eq:TMS_NS_t3} 
\end{IEEEeqnarray}
where \eqref{eq:TMS_NS_t1} is the power series representation of the operator exponential $ e^{\tau_1^*\hat{a}_R\hat{a}_I} $ and \eqref{eq:TMS_NS_m0} is from applying annihilation operators $\hat{a}_R\hat{a}_I$ $u$ times on the two-mode Fock state $\ket{km}$.
The upper limit on the sum in \eqref{eq:TMS_NS_m0} is because $\hat{a}\ket{0}=0$.
Furthermore, \eqref{eq:TMS_NS_m1} follows from the power series of exponential and Fock states being eigenstates of the photon number operator $\hat{a}^\dagger\hat{a}$, \eqref{eq:TMS_NS_t2} is the power series representation of the operator exponential $e^{-\tau_1 \hat{a}_R^\dagger\hat{a}_I^\dagger}$, and \eqref{eq:TMS_NS_t3} is from applying creation operators $u$ times.
By orthonormality of the Fock states, $\bra{st} S(\omega_1)\ket{km}$ is not zero only when
\begin{align}
    s &= k-u+l \text{~and~} t = m-u+l.\label{eq:orthonormality}
\end{align}
We eliminate the summation over $l$ in \eqref{eq:TMS_NS_t3} by solving for $l$ in \eqref{eq:orthonormality}.  Since \eqref{eq:orthonormality} also implies that $l-u=s-k=t-m$, rearranging the terms yields \eqref{eq:focksqueezing}.

\section*{Acknowledgment}
The authors benefited from discussions with Amit Ashok, Animesh Datta, Prajit Dhara, and Janis N\"{o}tzel, as well as comments from Gail Bash and the anonymous reviewers.
This material is based upon High Performance Computing (HPC) resources supported by the University of Arizona TRIF, UITS, and Research, Innovation, and Impact (RII) and maintained by the UArizona Research Technologies department.

\end{document}